\documentclass[sigplan,screen]{acmart}

\acmSubmissionID{2179}

\usepackage{subfig}
\usepackage[normalem]{ulem}
\usepackage{xspace}

\iffalse
    \newcommand{\revisioncr}[1]{\textcolor{blue}{#1}}
    \newcommand{\revisioncrs}[1]{\textcolor{red}{#1}}
    \usepackage[showboxes,overlay]{textpos}
    \TPMargin*{3pt}
    \textblockcolour{cyan}

    \newcommand{\revisionno}[3]{%
        \begin{textblock*}{14mm}(\csname revpos#2\endcsname,#3)
            {\centering \textcolor{blue}{\tt #1} \par}
        \end{textblock*}
    }
\else
    \newcommand{\revisioncr}[1]{#1}
    \newcommand{\revisioncrs}[1]{#1}
    \newcommand{\revisionno}[3]{}
\fi

\usepackage{tikz}
\usetikzlibrary{shapes.geometric,arrows.meta,positioning,arrows}
\newcommand*\circled[1]{\tikz[baseline=(char.base)]{
            \node[shape=circle,draw,inner sep=0.5pt] (char) {#1};}}

\usepackage[]{hyperref}
\usepackage{soul}
\usepackage{ctable}
\usepackage{booktabs}
\usepackage{longtable}
\usepackage{float}
\usepackage{lscape}
\usepackage[T1]{fontenc}
\usepackage{algcompatible}
\usepackage{amsmath} 
\usepackage{xcolor}

\usepackage{amsfonts}
\usepackage{pifont}
\usepackage{balance}
\usepackage{array}
\newcolumntype{P}[1]{>{\centering\arraybackslash}p{#1}}
\usepackage{listings}
\definecolor{keyword}{HTML}{2F7AFF}
\definecolor{identifier}{HTML}{000000}
\definecolor{commentstyle}{HTML}{5E5E5E}
\lstdefinelanguage{mylanguage}{
  keywords={True, False, return, switch, if, in, while, do, else, case, break, def, for},
  keywordstyle=\color{keyword}\bfseries,
  ndkeywords={class, export, boolean, throw, implements, import, this},
  ndkeywordstyle=\color{darkgray}\bfseries,
  identifierstyle=\color{identifier},
  sensitive=false,
  comment=[l]{\#},
  morecomment=[s]{/*}{*/},
  commentstyle=\color{commentstyle}\ttfamily,
  stringstyle=\color{red}\ttfamily,
  morestring=[b]',
  morestring=[b]"
}
\lstdefinestyle{mystyle}{
    language=mylanguage,
    backgroundcolor=\color{white},
    numberstyle=\tiny\color{black},
    basicstyle=\ttfamily\footnotesize,
    breakatwhitespace=false,
    breaklines=true,
    captionpos=b,
    keepspaces=true,
    numbers=left,
    numbersep=3pt,
    stepnumber=1,
    showspaces=false,
    showstringspaces=false,
    showtabs=false,
    tabsize=2,
    morekeywords={entry},
    frame=tb,
}
\newcommand{\compactparagraph}[1]{\noindent{\textbf{\textit{#1}}}}
\usepackage[ruled,vlined,linesnumbered]{algorithm2e}
\usepackage[normalem]{ulem}
\usepackage{multirow}
\usepackage{multicol}
\usepackage{makecell}
\usepackage{xspace}
\usepackage{enumitem}
\usepackage{lipsum}
\usepackage{ragged2e}
\usepackage{diagbox}
\usepackage{MnSymbol}
\usepackage{mathrsfs}
\definecolor{cell1}{HTML}{FF5B00}
\definecolor{cell2}{HTML}{ffa200}
\definecolor{cell3}{HTML}{f9da24}
\definecolor{cell4}{HTML}{f9da24}
\definecolor{cell5}{HTML}{cff09e}

\newcommand{\optype}[1]{\tt}

\usepackage{colortbl} 
\newcommand{\agrawal}[1]{\textcolor{red}{``GA: #1''}}
\long\def\agrawal#1{\textcolor{red}{``GA: #1''}}
\long\def\wniu#1{\textcolor{blue}{``WN: #1''}}
\long\def\hidetext#1{}

\copyrightyear{2026}
\acmYear{2026}
\setcopyright{cc}
\setcctype{by}
\acmConference[ASPLOS '26]{Proceedings of the 31st ACM International Conference on Architectural Support for Programming Languages and Operating Systems, Volume 2}{March 22--26, 2026}{Pittsburgh, PA, USA}
\acmBooktitle{Proceedings of the 31st ACM International Conference on Architectural Support for Programming Languages and Operating Systems, Volume 2 (ASPLOS '26), March 22--26, 2026, Pittsburgh, PA, USA}
\acmPrice{}
\acmDOI{10.1145/3779212.3790164}
\acmISBN{979-8-4007-2359-9/2026/03}

\begin{document}
\pagenumbering{gobble}

\newcommand{\projectnamenott}{FlashMem}
\newcommand{\projectname}{{\tt \projectnamenott}\xspace}
\newcommand\nott[1]{\bgroup\let\tt\relax#1\egroup}

\title{\projectnamenott: Supporting Modern DNN Workloads on Mobile with GPU Memory Hierarchy Optimizations}

\author{Zhihao Shu}
\email{Zhihao.Shu@uga.edu}
\affiliation{%
  \institution{University of Georgia}
  \city{Athens}
  \state{GA}
  \country{USA}
}

\author{Md Musfiqur Rahman Sanim}
\email{musfiqur.sanim@uga.edu}
\affiliation{%
  \institution{University of Georgia}
  \city{Athens}
  \state{GA}
  \country{USA}
}

\author{Hangyu Zheng}
\email{hyzheng@uga.edu}
\affiliation{%
  \institution{University of Georgia}
  \city{Athens}
  \state{GA}
  \country{USA}
}

\author{Kunxiong Zhu}
\email{Kunxiong.Zhu@uga.edu}
\affiliation{%
  \institution{University of Georgia}
  \city{Athens}
  \state{GA}
  \country{USA}
}

\author{Miao Yin}
\email{miao.yin@uta.edu}
\affiliation{%
  \institution{University of Texas at Arlington}
  \city{Arlington}
  \state{TX}
  \country{USA}
}

\author{Gagan Agrawal}
\email{gagrawal@uga.edu}
\affiliation{%
  \institution{University of Georgia}
  \city{Athens}
  \state{GA}
  \country{USA}
}

\author{Wei Niu}
\email{wniu@uga.edu}
\affiliation{%
  \institution{University of Georgia}
  \city{Athens}
  \state{GA}
  \country{USA}
}

\renewcommand{\shortauthors}{Zhihao Shu et al.}
\renewcommand{\shorttitle}{\nott{\projectname}}

\begin{abstract}
The increasing size and complexity of modern deep neural networks (DNNs) pose significant challenges for on-device inference on mobile GPUs,  with limited  memory and computational resources. 
Existing DNN acceleration frameworks  primarily deploy a  {\em weight preloading}  strategy, 
where all model parameters are loaded into memory before execution on mobile GPUs. 
We posit that this approach is 
not adequate for modern DNN workloads that comprise very large model(s) and possibly execution 
of several distinct models in succession. 
In this work, we introduce FlashMem, a  memory streaming framework designed to efficiently execute {\em large-scale modern DNNs} and {\em multi-DNN} workloads  while minimizing memory consumption and reducing inference latency.
Instead of fully preloading weights, FlashMem statically determines model loading schedules and dynamically streams them on demand, leveraging 2.5D texture memory to minimize data transformations and improve execution efficiency. 
Experimental results on 11 models demonstrate that FlashMem achieves 2.0$\times$ to 8.4$\times$ memory reduction and 1.7$\times$ to 75.0$\times$ speedup compared to existing frameworks, enabling efficient execution of large-scale models and multi-DNN support on resource-constrained mobile GPUs.
\end{abstract}


  
\maketitle

\section{Introduction}

Deep Neural Networks (DNNs) have become an integral part and core enabler of modern AI applications, spanning from virtual assistants~\cite{vu2024gptvoicetasker, taneja2024jill, guan2023intelligent} to real-time healthcare diagnostics~\cite{saeed2024predictive, miao2022real}.
The demand for on-device DNN  execution directly on mobile devices has surged due to the growing concerns over privacy, the need for real-time responsiveness, and the limitations of cloud-dependent processing~\cite{li2024flexnn,guo2023sti}. 

AI-powered mobile applications often rely on multiple models executing in a short span~\cite{jeong2022band,wang2022stitching} --  for example, a real-time image processing pipeline that involves encoder, detection, and segmentation. This is necessitating support for {\em multi-DNN} 
execution. In parallel, 
individual model has continued to scale in size and complexity~\cite{llama2, StableDiffusion,niu2024sod2}.
In summary, with emerging application needs, 
 mobile DNN frameworks need to be able to efficiently manage the execution of multiple models, each with substantial (and growing) memory and computation requirements.  

Contrary to \revisioncrs{these requirements}, current state-of-the-art mobile DNN frameworks~\cite{TensorFlow-Lite,executorch,llamacpp,Ni_ncnn_2017,chen2018tvm,Ali-MNN,llamafile,ollama,niu2020patdnn,niu2022gcd} use a preloading strategy that loads the entire model -- its weights and computational graph -- into memory before starting inference.
While this approach improves inference speed by reducing storage access latency, 
it also significantly increases peak memory consumption, straining the limited memory resources of mobile hardware~\cite{guo2023sti}. 
Table~\ref{tab:memory_size_usage} presents the memory footprint of different DNNs when executed on a OnePlus 12 mobile device using the MNN~\cite{Ali-MNN} framework. Even for individual models, the memory consumption is substantial, making it almost impossible  to support multi-model execution without exceeding system limitations.

Beyond memory constraints, mobile GPUs present additional performance bottlenecks due to their unified memory architecture, where the CPU and GPU share memory but operate in distinct execution spaces. 
Unlike desktop-class GPUs with dedicated VRAM, mobile GPUs rely on specialized texture memory (and cache) to optimize data access~\cite{liang2022romou}. 
Current frameworks perform weight transformations at the computational graph level, leading to redundant data transformations ~\cite{niu2024smartmem},
excessive GPU kernel launches, and inefficient memory management. 
These inefficiencies are significantly compounded in multi-DNN workloads, where models frequently swap in and out of memory to accommodate different tasks.
As also shown in Table~\ref{tab:memory_size_usage}, GPU initialization overhead (including 
the data transformation time) can  easily dominate 
the inference latency. 

To address these challenges, we propose \projectname, 
a memory streaming framework that leverages hierarchical GPU memory optimizations to efficiently support {\it modern DNN} workloads and {\it multi-DNN} workloads on mobile devices.
\projectname both minimizes memory consumption and reduces  inference latency.
Unlike existing frameworks that preload entire models into memory, \projectname streams weights dynamically during inference to keep memory usage within device constraints.
Furthermore, it  improves data transformations efficiency by leveraging the GPU's texture memory to process weights efficiently, while  also overlapping  weight loading with computation for better I/O efficiency.

The detailed contributions of \projectname are as follows:
\begin{itemize}[leftmargin=*,noitemsep,nolistsep]
    \item {\it Optimized Overlap Plan Generation (OPG)}: we formalize the OPG problem, \revisioncrs{which defines when and where each weight should be loaded, } 
    and develop LC-OPG, a load-aware solver that minimizes peak memory usage while ensuring efficient execution.
    \item {\it Load Capacity Profiling and Adaptive Fusion}: we classify operators by memory and compute intensity, to determine the load capacity and an adaptive fusion mechanism to balance memory and performance.
    \item {\it Hierarchical GPU Memory Optimization}: we optimize 2.5D texture memory by restructuring weight layouts for efficient caching, reducing transformation overhead, and improving memory efficiency.
    \item {\it Pipeline-Aware Kernel Execution}: we introduce a branch-free, pipelined execution strategy that interleaves computation with weight loading, maximizing GPU efficiency for large-scale DNN inference.
\end{itemize}

Compared to other frameworks such as ExecuTorch~\cite{executorch}, MNN~\cite{Ali-MNN}, NCNN~\cite{Ni_ncnn_2017}, LiteRT (formerly TensorFlow-Lite)~\cite{TensorFlow-Lite}, TVM~\cite{chen2018tvm,mlc-llm},  
and SmartMem~\cite{niu2024smartmem}, \projectname supports overlapping model loading and model 
execution, optimizing both memory and computation, which enables modern large-scale DNN and 
multi-model execution. The results demonstrate that \projectname reduces average memory usage by 2.0 $\times$ to 8.4$\times$ memory reduction \revisioncrs{on all evaluated models}, and up to $10.1\times$ on certain  models like DeepViT.
Compared to SmartMem, \projectname achieves an average speedup of 8.6$\times$, with up to 15.8$\times$ acceleration for GPT-Neo 1.3B and 9.3$\times$ for SD-UNet, while maintaining a better trade-off between memory efficiency and inference latency.
\projectname also achieves 1.7$\times$ to 75.0$\times$ speedups compared to all state-of-the-art methods.

\begin{table}[t!]
    \centering
    \footnotesize
    \setlength{\tabcolsep}{4pt}
    \caption{Memory usage and latency of various models on the OnePlus 12. ``Avg.'' indicates average memory consumption; ``Trans.'' represents data transformation latency.}
    \begin{tabular}{lc|cc|ccc}
        \toprule \multirow{2}{*}{Model}  & \multirowcell{2}{\# Parameters\\ (M)} & \multicolumn{2}{c|}{Memory (MB)} & \multicolumn{3}{c}{Latency (ms)} \\
        ~                                & ~                                     & Peak                             & Avg.                            & Load  & Trans. & Infer \\
        \hline
        Whisper~\cite{radford2023robust} & 356                                   & 4,077                            & 1,650                           & 2,702 & 3,441  & 1,343 \\
        GPTNeo~\cite{gpt-neo}            & 164                                   & 1,026                            & 610                             & 631   & 2,898  & 337   \\
        SD-UNet~\cite{StableDiffusion}   & 860                                   & 4,858                            & 1,800                           & 4,159 & 17,588 & 1,647 \\
        \bottomrule                       
    \end{tabular}
    \label{tab:memory_size_usage}
\end{table}

\section{Background and Motivation} 
\label{sec:background}

\subsection{GPU Memory Hierarchy on Mobile Platforms} 
\begin{figure}[t!]
  \centering
  \includegraphics[width=1\columnwidth]{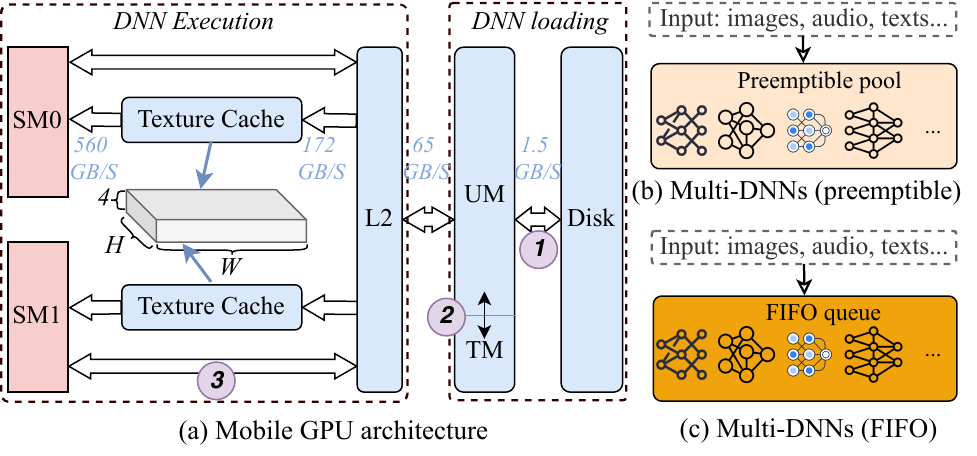}
  \caption{Left (a) shows the hierarchical memory structure and bandwidth on
  mobile GPUs, highlighting the multi-step weight transfer path from disk to streaming
  multiprocessors. Right (b and c) depicts two multi-DNN support strategies.}
  \label{background:texture_memory_gpu}
\end{figure}

Mobile GPUs typically use a hierarchical memory organization that includes unified memory (UM) and texture memory (TM).
Given the significant resource requirements of executing DNNs, optimizing for this memory 
hierarchy is important. 
As a background, the size of a deep neural network is largely determined by its weight tensors, which consume most of the model's storage footprint. 
Figure~\ref{background:texture_memory_gpu} (a) illustrates the multi-step path to transfer these weights from disk to UM (\circled{1}), moving through UM to TM (\circled{2}), and executing on SM (streaming multiprocessor\circled{3}).
Modern mobile GPUs (e.g., ARM Mali and Qualcomm Adreno) overlap compute with memory transfers by using independent command queues.

The texture memory subsystem is specifically designed for a two-dimensional layout with a dedicated cache to exploit 2D spatial locality. 
This specialized memory organization is known as 2.5D texture memory.
Unlike conventional one-dimensional memory, which flattens tensor dimensions into a linear sequence, the 2.5D approach reorganizes multi-dimensional tensors into small two-dimensional tiles with limited depth -- typically four channels or scalar elements (see Figure~\ref{background:texture_memory_gpu} (a)).
Romou~\cite{liang2022romou} has demonstrated that texture memory can accelerate DNN execution by up to 3.5$\times$ speedup over unified-memory-based approaches. 
\revisionno{Review-B\\(1d)}{R}{0.0cm}
\revisioncr{SmartMem~\cite{niu2024smartmem} optimizes 2.5D memory layouts by systematically eliminating costly layout transformations (e.g., {\tt  Reshape}, {\tt Transpose}). This approach has proven particularly effective for Transformer architectures on mobile GPUs, yielding significant speedups in inference. } 
Despite these gains, loading the weights from disk into the GPU texture  still involves 
inefficiency of intermediate copies, each step incurring overhead and memory demands~\cite{jia2022codl,liang2022romou}.

\subsection{Demands for Multi-model Support on Mobile} 
Mobile workloads often execute multiple models~\cite{Xiang2019pipeline,jeong2022band}, rather than relying on a single DNN that is persistently stored  and executed. 
As shown in Figure~\ref{background:texture_memory_gpu} (b) and (c), 
two representative approaches for multi-DNN usage on mobile devices are  \emph{FIFO} ~\cite{Xiang2019pipeline,wang2022stitching} and \emph{preemptive}~\cite{jeong2022band,han2024pantheon} scheduling. FIFO scheduling employs a sequential design where multiple models are queued and executed sequentially until all requests are fulfilled.
Preemptive solutions permit a high-priority model to interrupt or replace a lower-priority model during execution, which can be valuable for time-sensitive applications, but are not the focus of 
our study. 


Some of the concrete motivating scenarios for FIFO-style execution are as follows. 
In camera-based augmented reality, 
an object detection DNN (e.g., ResNet~\cite{resnet}) might run briefly to identify key objects in the scene, then a separate model checks for user actions~\cite{gpt-neo} or  performs depth analysis~\cite{depth_anything_v2}, each triggered only occasionally. 
Another scenario appears in user-facing translators that chain small, specialized models -- such as a speech recognition~\cite{radford2023robust} model followed by other actions with the recognized text, e.g., image generation~\cite{StableDiffusion}.

\begin{figure}[t!]
  \centering
  \includegraphics[width=0.95\columnwidth]{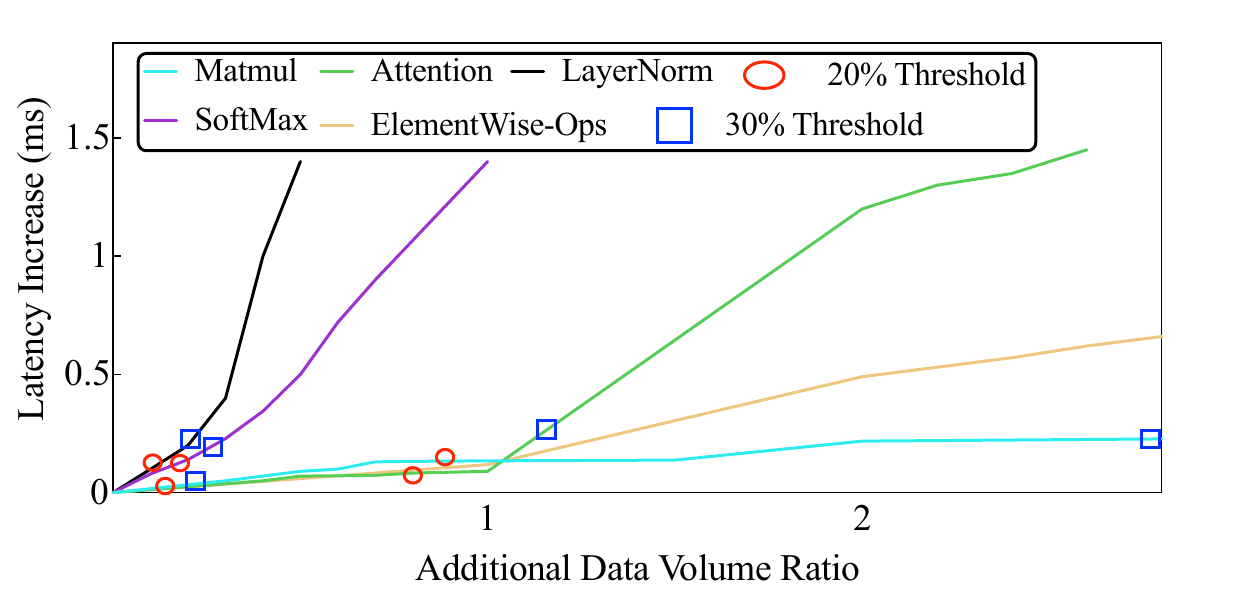}
   \caption{\revisionno{Review-A\\(3)}{L}{0.0cm}\revisioncr{Latency increase (ms) across different operators on mobile GPUs due to additional data transformations. The X-axis shows the ratio of additional data loaded compared to the original kernel input (e.g., 1.0 indicates an equal amount of extra data is streamed alongside computation). Threshold markers show where latency overhead reaches 20\% and 30\% of the original kernel.}}

\label{motiv:load_capacity}
\end{figure}

In each of the above scenarios, we face the following challenges. 
{\it Storing all models fully in GPU memory simultaneously would be infeasible for large sets (or large-scale) of DNNs. }
Executing them one-by-one via a simple FIFO strategy, on the other hand,  avoids memory contention but has 
large overhead, since each model  needs to load from disk and convert its weights into texture-based layouts before execution. 
{\it This paper addresses such FIFO multi-DNN pipelines by mitigating repeated load and layout transition 
overhead, providing faster inference across a series of distinct  models,  none of which is invoked 
for inference a large number of times in succession. }

\subsection{An Experiment  Motivating Our Design} 
\label{sec:motiv}

Overlapping data movement with kernel execution not only hides transfer latency but can 
also help limit the overall memory footprint. 
Staging weights or intermediate tensors in small increments and streaming them into GPU texture memory as needed helps reduce peak memory usage in multi-model pipelines.
However, effectively interleaving data transformations with kernel arithmetic requires an understanding of how different kernels respond to an overlapping operation. 
Figure~\ref{motiv:load_capacity} shows a preliminary study that measures the increase in latency of each kernel when forced to perform additional data transfers in parallel. 
{\tt Softmax} and {\tt LayerNorm}, for instance, exhibit substantial slowdowns even at modest transfer volumes, indicating limited  opportunities for effective overlap. 
Element-wise operators (e.g., {\tt Activation}, {\tt Add}) allow larger data streams before latencies approach the overhead of a dedicated copy call, 
while {\tt Matmul} tolerates more significant data inflow with relatively small  latency increases. 
This experiment shows that a  careful formulation and solution  is needed to effectively overlap 
data movement  with computations.

\section{Generating a Data Movement Schedule }
\label{sec:schedule-definition}

\subsection{Problem Formalization of OPG}
\label{sec:cp_sat_overlap}

We first formalize the Overlap Plan Generation (OPG) problem
 and
\revisioncrs{reduce it to the Constraint Programming Satisfiability (CP-SAT)~\cite{cpsatlp} problem}.
\revisionno{Review-B\\(1d)}{R}{0cm} 
\revisioncr{Later, we integrate Google OR-Tools~\cite{cpsatlp} -- an open-source software suite for combinatorial optimization, to solve our overlap planning problem.} 

A Deep Neural Network (DNN) is represented by a directed acyclic graph (DAG) $G=(V,E)$, where each node 
(operator or layer, used interchangeably) $v \in V$ consumes and produces tensors or {\em weights}.  An edge $(u,v)\in E$ indicates that the output  weight of layer $u$ is used by the layer $v$. 
The actual order of execution of the operators (determined in some fashion outside the scope of our work)  imposes a linear order of execution of  
 $\{\,1,2,\dots,N\}$. For simplicity of presentation, it is assumed that each node produces exactly one weight or tensor, which can also be numbered  $\{\,1,2,\dots,N\}$.
 Let $i_w$ denote the earlier layer that consumes the weight $w$.  We actually consider two 
 separate problems, driven by the need to reduce the use of unified memory, and for addressing 
 a hard limit on the amount of texture memory.
\revisionno{Review-A\\(4)}{R}{0.0cm} 
\revisioncr{The key notations and symbols we used in this section are summarized in Table~\ref{tab:math-symbols}.}

\begin{table}[t]
    \centering
    \small
    \setlength{\tabcolsep}{8pt}
    \caption{\revisionno{Review-A\\(4)}{L}{0.0cm}\revisioncr{Summary of the key notations and symbols.}}
    \label{tab:math-symbols}
    \begin{tabular}{l l}
        \toprule
        \textbf{Symbol} & \textbf{Description} \\
        \midrule
        $W$ & Weights to preload \\
        $z_w$ & Earliest layer loading $w$ to unified memory \\
        $x_{w,\ell}$ & Chunks of $w$ transformed at layer $\ell$ \\
        $i_w$ & Layer index consuming weight $w$ \\
        \midrule
        $C_\ell$ & Load Capacity (max load at layer $\ell$) \\
        $M_{peak}$ & Peak memory limit for transformations \\
        $T(w)$ & Total chunks for weight $w$ \\
        \midrule
        $\lambda, \mu$ & Penalty terms (preload/distance) \\
        $\alpha$ & Threshold for splitting fused operators \\
        \bottomrule
    \end{tabular}
\end{table}

 \subsubsection{Unified Memory Usage }

Focusing on unified memory usage, the OPG {\em decision variables} are  as follows:
\begin{itemize}[leftmargin=1.5em,noitemsep,nolistsep]
\item $W$: The  set of weights  that will be loaded from disk to unified memory and transformed by specialized data-loading kernels before the actual DNN execution. 
\\
\textbf{Interpretation:} Usually the  first few operators do not have preceding layers to load and transform their weights, so their weight tensors must be in the set $W$. Thus,  to 
meet all constraints, we can selectively put additional weights into the set $W$. However, additional weights in the set $W$  increase the data loading time for the model and potentially the memory usage 
as well during the execution of initial layers. 
\item $z_{w} \in \{1,2,\dots,N\}$: Index of the  earliest layer that loads the  weight $w$ from disk to unified memory.
    \\
    \textbf{Interpretation:} This set of decision variables is associated with 
    {\em loading distance}, which is the  difference $i_w - z_{w}$ that  captures how many layers in advance the weight $w$ is loaded into unified memory. Note that once a layer initiates loading weights, these weights must be fully loaded from disk and remain in unified memory until their last use.  
\emph{If $i_w - z_w$ is small,} weight $w$ is loaded closer to its first consumption, reducing memory footprint but potentially risking concurrency with compute-intensive steps.  
\emph{If $i_w - z_w$ is large,} weight $w$ is loaded earlier, which might mitigate concurrency for consumption but raises the overall memory footprint due to prolonged residency.
Including the loading distance in our objective function and constraints allows us to explicitly balance timely data availability with the total  memory footprint. 

\end{itemize}

Our solver chooses the above values subject to our objective function. 

\compactparagraph{Objective Function.}
We minimize:
\[
   \lambda \times W
   \;+\;
   (1 - \lambda) \sum_{w} \bigl(i_w - z_w\bigr)
\]
where:
\begin{itemize}[leftmargin=1.5em,noitemsep,nolistsep]
\item The total memory usage of the preloaded weights  before inference is reflected in $W$. 
\item $(i_w- z_w)$ is the loading distance for weight $w$ - the corresponding term  penalizes extended residency.
\item $\lambda$  is a term  that establishes  the weight  of   preload overhead and loading distance. 

\end{itemize} 

\subsubsection{Texture Memory Constraints} 

 For a fine-grained control of the weights transforming process (from unified memory to texture memory), we split each weight $w$  into chunks with a uniform chunk size of $S$. Let $T(w)$ denote the number of chunks of size $S$ that the weight $w$ is partitioned into.  Now, the decision variable we have 
 here is  $x_{w,\ell} \in \{0,1,\dots,T(w)\}$: the number of chunks in weight $w$ that will be transformed (from unified memory to texture memory) by the layer $\ell$.   Larger values for the  variable $x_{w,\ell}$ for 
 initial layers imply that  more data is transformed early, potentially increasing memory usage, but also ensuring timely data availability. Conversely, smaller values preserve memory but risk idle time  with later computations. 

 \compactparagraph{Constraints.}
Let \(L(w)\) represent the set of layer indices that load at least one chunk of weight \( w \). Each layer $\ell \in L(w)$ has the  constraint:
\[
L(w) = \{\,n, n+1, \dots, m \mid m \geq n > 0,\; i_w > m \geq 1\},
\]
 Note that the indices from \( n \) to \( m \) are \emph{not required to be continuous}; layers within this range can be selectively chosen based on scheduling constraints. Each selected layer index falls within the range \(\{1, 2, \dots, i_w - 1\}\).

\noindent\emph{{\bf (C0)} Completeness of Allocation:}
\[
\sum_{\ell \in L(w)} x_{w,\ell} = T(w),
\quad \forall w.
\]

\noindent\emph{{\bf (C1)} Loading Distance Implication:}
\[
\bigl(x_{w,\ell} \ge 1\bigr)\; \Rightarrow\; (z_w \le \ell),
\quad \forall w,\forall \ell \in L(w).
\]
If the weight $w$ is partially assigned to the layer $\ell$, the earliest load index cannot exceed $\ell$.

\noindent\emph{{\bf (C2)} Layer Transformation  Memory Limit:} 
$M_{\mathrm{peak}}$ is an integer variable that limits 
\revisionno{Review-B\\(1c)}{R}{0.0cm} \revisioncr{the total in-flight memory (spanning weights in both unified and texture memory) allowed during execution;} 
and as stated earlier, $S$ is the size of one chunk. 

\[
\underbrace{\sum_{w\,:\,I(w)>\ell} x_{w,\ell} \times S}_{\text{Total chunks memory transformed by layer l}} \le M_{\mathrm{peak}},
\quad \forall \ell.
\]

\compactparagraph{CP-SAT Reduction.}
By mapping the chunk allocations $(x_{w,\ell})$ and earliest load indices $(z_w)$ to integer variables, along with logical constraints (e.g., $(x_{w,\ell}\ge1) \Rightarrow (z_w\le\ell)$), the OPG problem \revisioncrs{can be reduced} to the  well-known existing optimization problem: CP-SAT.  

\subsection{Load Capacity Aware OPG-Solver}
\label{sec:op_categories}

We introduce a Load Capacity ($C_\ell$) Aware Overlap Plan Generation (LC-OPG) Solver, with tailored constraints for weight loading and transformation scheduling. 
\revisionno{Review-B\\(1a, 1b)}{R}{0cm}
\revisioncr{$C_\ell$ represents the maximum number of byte chunks that layer $\ell$ can concurrently transform from unified memory to texture memory without incurring significant overhead.} 
For a general DNN DAG, we  can define the CP-SAT model in Google OR-Tools\revisioncr{~\cite{cpsatlp}} as follows. 
First, we create integer variables for each $x_{w,\ell}$ and $z_w$, put  the weights of the first layer in $W$ and enforce  constraints (0)–(2). If any layer violates the constraint $(C2)$ by chunk allocations, a fallback mechanism puts the weight, which takes the most memory that will be loaded by this layer, to $W$, then re-runs until generating the full plan.  
Upon convergence, 
the resulting allocations specify the layer-weight preload assignments, including a mapping that specifies which weight segments will be preloaded from unified memory to \revisioncrs{GPU} texture memory, along with their corresponding start and end offsets. 
\revisionno{Common\\(1)\\Review-D\\(1)}{L}{0.0cm}
\revisioncr{The solver targets the memory lifetime directly and runs offline for each model prior to deployment, generating a reusable overlap plan that incurs no runtime overhead during inference.}
This overlap plan serves as a guide for runtime scheduling, enabling the effective overlap of data transformations with computation, which in turn minimizes the memory footprint while maintaining performance.


However, the CP-SAT solver~\cite{cpsatlp} operates in an explosive combinatorial search space, making naive scheduling infeasible within practical runtime limits.
To reduce search complexity, we introduce a set of structured constraints:

\noindent\emph{{\bf (C3)} Layer Load Capacity Constraint}
\begin{equation}
    \sum_{w \in W_\ell} x_{w,\ell} \leq C_\ell, \quad \forall \ell \in \{1,\dots,N\}
\end{equation}
Prevents exceeding the predicted \textbf{$C_\ell$} for each layer.

\noindent\emph{{\bf (C4)} Dynamic Load Thresholding}
If layer $\ell$ exceeds its capacity $C_\ell$, we apply a \textbf{fallback strategy}:
\begin{itemize}[leftmargin=1.5em,noitemsep,nolistsep]
    \item \textbf{Soft Thresholding:} Dynamically adjusts $C_\ell$ based on available memory.
    \item \textbf{Weight Prioritization:} Prefers to load weights closer to execution while delaying others.
\end{itemize} 

\begin{table}[t!]
    \setlength{\tabcolsep}{2pt}
    \centering
    \small
    \caption{Comparison of Google CP-SAT and Our LC-OPG Solver: ``A.P.M.C.'' for
    Adaptive Peak Memory Control, ``P.L.D.S.'' for Per-Layer Dynamic Scheduling, and ``M.O.C.''
    for Memory, Overlap, and Compute Load.}
    \label{tab:solver_comparison}
    \begin{tabular}{lcc}
        \toprule Feature            & Google CP-SAT & Our LC-OPG Solver  \\
        \midrule Memory Constraints & Static limits & A.P.M.C.            \\
        Load Capacity Awareness     & Not included  & P.L.D.S.            \\
        Solver Convergence          & May fail      & Fallback mechanism \\
        Objective Function          & Generic       & Balances M.O.C.     \\
        \bottomrule
    \end{tabular}
\end{table}

Elaborating on C4, if the CP-SAT solver fails to find a feasible schedule due to strict constraints or memory overuse, we incorporate a structured fallback strategy to ensure convergence of the solution . First, a \textit{soft threshold adjustment} dynamically relaxes the load capacity constraint \( C_\ell \), allowing minor deviations to accommodate weight transformations that are critical for execution. If a feasible schedule remains unreachable, the solver applies \textit{incremental preloading}, where selected weight tensors are preloaded ahead of time to reduce scheduling complexity. Finally, if the problem persists, the solver switches to a \textit{greedy heuristic backup}, which prioritizes memory-efficient weight allocations while maintaining overlap as much as possible. This tiered fallback mechanism prevents excessive computational overhead and guarantees a solution within practical runtime limits.

Table~\ref{tab:solver_comparison} compares our updated solver aided by  our fallback strategy and per-layer scheduling with the original Google CP-SAT solver. The determination of the load capacity will be 
discussed in the next subsection.


\compactparagraph{Implementation Considerations.}
Our solver is implemented using {Google OR-Tools} \cite{cpsatlp} with multiple optimizations tailored to minimize search complexity and execution overhead. 
To prevent excessive memory footprints, we employ an {\em incremental scheduling} approach, which processes weight transformations in a rolling window. 
\revisionno{Review-A\\(5)}{R}{0.0cm}
\revisioncr{This mechanism gradually builds the overlap plan by updating scheduling decisions as new layers in the computational graph activate and earlier ones complete. It maintains a manageable number of active constraints, ensuring predictable solver runtime.} 

Additionally, the solver incorporates {profile-guided adjustments}, dynamically updating \( C_\ell \) based on real-time profiling of execution behavior and available memory. This adaptive approach enables more informed decision making and mitigates the risk of scheduling conflicts. Lastly, we introduce a {hybrid execution mode} that seamlessly switches between CP-SAT and heuristic scheduling based on workload conditions. 
If the solver returns an infeasible solution or only finds a feasible (but not optimal) solution after exceeding the time limit, the system falls back to heuristic strategies to ensure timely execution. 
These fallback mechanisms enhance the robustness and scalability of the LC-OPG solver for real-world DNN deployment.  

\begin{table}[t!]
    \centering
    \small
    \caption{\revisioncr{Execution time breakdown for the evaluated models (constrained by a 150-second solver limit).}}
    \begin{tabular}{lrrrl}
        \toprule
        Model & Process & Build CP-SAT & Solve & Solver \\
              & nodes (s) & model (s) & model (s) & Status \\
        \midrule
        GPTN-S      & 0.010 & 0.260  & 45.00  & OPTIMAL \\
        GPTN-1.3B   & 0.020 & 1.170  & 121.00 & FEASIBLE \\
        GPTN-2.7B   & 0.050 & 1.980  & 121.00 & FEASIBLE \\
        ViT-8B      & 0.001 & 4.110  & 121.40 & FEASIBLE \\
        Llama2-13B  & 0.007 & 3.566  & 124.80 & FEASIBLE \\
        Llama2-70B  & 0.023 & 14.456 & 136.38 & FEASIBLE \\
        \bottomrule
    \end{tabular}
    \label{tab:solver-costs}
\end{table}

\revisionno{Common\\(2)\\Review-D\\(2)}{R}{0.0cm}
\revisioncr{
To validate convergence, we evaluate the solver for different models on a workstation with 512 GB of DRAM and an AMD 5995WX CPU (128 threads).
Table \ref{tab:solver-costs} presents a runtime comparison.
We empirically set a 150-second time limit to balance solution quality and analysis cost, 
as LC-OPG consistently converges to feasible or near-optimal plans within this timeframe.
A potential corner case we observed is for dynamic neural networks, 
where runtime-dependent execution paths can increase solver time due to the need to explore multiple possible execution branches. 
We leave this as a future work since it is out of the scope in this paper.
It is important to note that the runtime cost increases non-linearly with model size. This is due to the solver's constraints, which prune infeasible regions using load-capacity thresholds and incremental scheduling.
}


\compactparagraph{Hyperparameters Considerations And Adaptivity.}
In order to achieve the optimal performance under defined memory limits, the solver must carefully balance runtime memory during inference and memory usage at initialization.  $M_{peak}$, introduced earlier, 
 sets an upper bound of how much memory we can use at any DNN layer during inference, 
 but  does not include the memory used by the persistent weights that have already loaded and transformed at initialization.
 Therefore, the combined memory usage of the persistent weights \( W \) and \( M_{peak} \) defines our peak and stable memory footprint.
 
Our hyperparameters also allow for prioritizing memory usage over execution time or vice-versa.  
To reduce memory requirements, we should preload as few weights as possible, since overlapping gives us a fine-grained control of weights instead of loading them all at once. On the other hand, if the goal is to reduce 
the latency for execution (after start of execution),  we should preload as many as possible, since it reduces workload for each kernel. A higher preload ratio can be achieved by increasing $M_{peak}$ -- this also 
reduces the problem complexity so that the solver can easily  converge. On the other hand, 
to lower the preload ratio, we should set smaller $M_{peak}$. Empirically, for memory priority, we set  $M_{peak}$ as 500 MB, where $\lambda$ is set close to 0.9.

\section{Design of \projectnamenott}
\label{sec:design-details}

\subsection{System Overview of \projectnamenott}
\label{sec:overview}

\begin{figure}[t!]
  \centering
  \includegraphics[width=0.95\columnwidth]{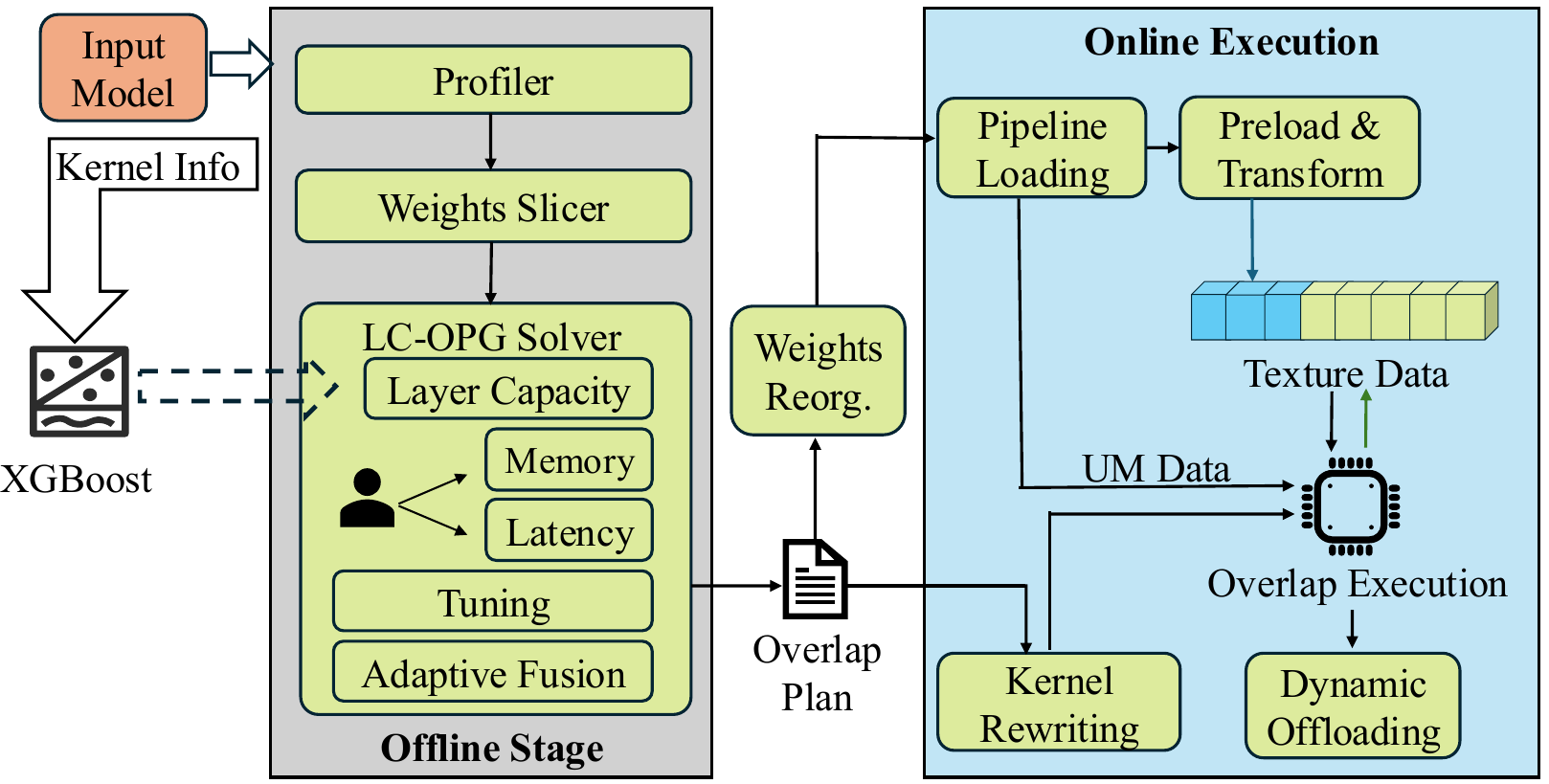}
  \caption{\textbf{Overview of \projectname.} }
  \label{design:system_overview}
\end{figure}

\projectname builds on an existing end-to-end DNN execution framework SmartMem~\cite{niu2024smartmem}, leveraging its operator-level layout transformations while adding specialized modules for overlap planning,  kernel rewriting, and adaptive fusion.  Figure~\ref{design:system_overview} illustrates the overall workflow.
We begin by parsing the DNN model to identify each layer's structure and then use capacity-based predictions (Section~\ref{subsec:loadcapacity}) to estimate how much additional data each layer can load without causing significant slowdown. 
The  CP-SAT solver described in previous section  then produces an \emph{overlap plan} that schedules weight-loading tasks. 
\revisionno{Review-B\\(1f)}{L}{0.0cm} 
\revisioncr{When global constraints (i.e., C0--C4) fail,}  
certain large fused operators may be selectively unfused to ensure sufficient load capacity (Section~\ref{sec:weight_mnager}). 
Next, with the overlap plan in place, we rewrite each GPU kernel (Section~\ref{sec:kernel_rewriting}) to embed data loading operations within its computation -- the goal is to 
 interleave  arithmetic operations with continuous, vectorized data reads, maximizing GPU utilization and avoiding branch divergence.


\subsection{Determining  Load Capacity}  
\label{subsec:loadcapacity}

To assess the load capacity of various operators, we first classify them based on their memory bandwidth usage, tolerance for load capacity, and computational intensity.
We identify three types of operations as summarized in Table~\ref{tab:operator_classification}
\begin{itemize}[leftmargin=*, noitemsep, nolistsep] 
    \item \emph{Elemental operators} (\emph{e.g.,} {\tt Elementwise}, {\tt Activation}) typically rely on linear memory accesses and minimal internal dependencies. They are often memory-bound but allow for moderate concurrency in terms of data-loading tasks, given their simpler arithmetic footprint. 
    \item \emph{Reusable operators} (\emph{e.g.,} {\tt Conv}, {\tt MatMul}) exhibit structured data reuse and multi-dimensional computational patterns (e.g., loop tiling). They can accommodate high overlaps in data loading, while their higher arithmetic complexity necessitates careful resource allocation.
    \item \emph{Hierarchical operators} (\emph{e.g.,} {\tt Softmax}, {\tt LayerNorm}) involve intricate, stepwise computations and synchronizations, leaving limited bandwidth for concurrent data movement. 
\end{itemize}

\begin{figure}[t!]
    \begin{center}
        \centerline{\includegraphics[width=\columnwidth]{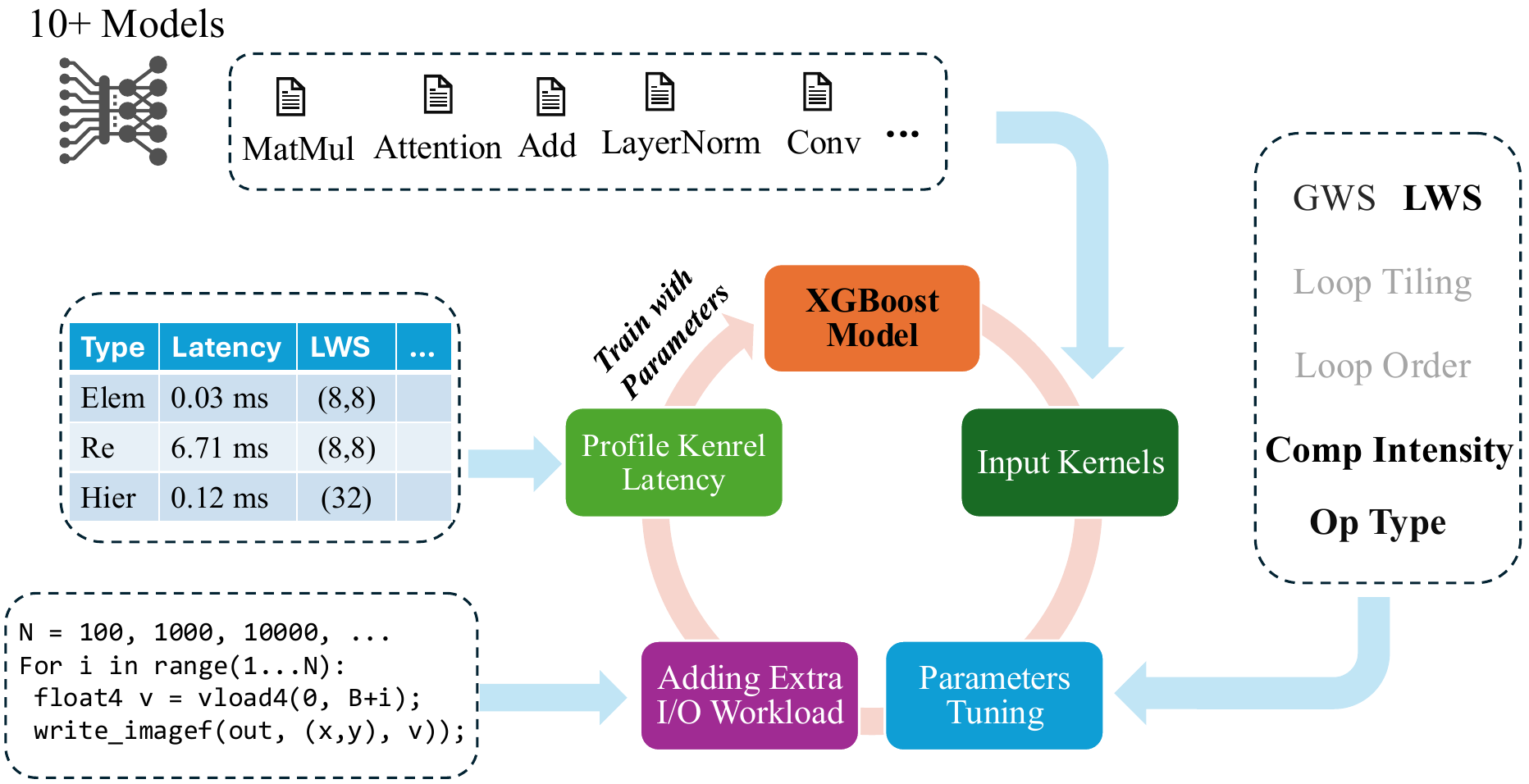}}
        \caption{Profiling operators from more than ten models. We systematically
        adjust key parameters in our framework to measure kernel latency changes.
        The collected data is then used to train our XGBoost-based regression
        model for accurate latency prediction. 
        \revisionno{Review-B\\(1e)}{R}{0.0cm} 
        \revisioncr{Global Work Size (GWS) and Local Work Size (LWS) determine workload distribution and thread partition, respectively. }}
        \label{design:xgboost} \vskip -0.5 cm
    \end{center}
\end{figure}

\vskip 0.2cm
\noindent In our work we  adopt a lightweight profiling-based method that strategically samples execution scenarios to estimate load capacity with minimal overhead.
We train an XGBoost regression model \cite{chen2016xgboost} to predict execution latency under varying additional loads (as shown in Figure~\ref{design:xgboost}).  
The model guides our CP-SAT solver (Section~\ref{sec:cp_sat_overlap}) by determining per-layer load capacity thresholds that ensure minimal performance degradation. Specifically, we set distinct thresholds based on operator sensitivity and baseline latency characteristics: 0\% for hierarchical operators, 20\% for reusable operators, and 300\% for elemental operators. 
Hierarchical operators are particularly sensitive to additional data loading and have relatively high baseline kernel latencies; thus we do not use this type of OPs to perform overlapping. Elemental operators, however, exhibit minimal latency growth under additional loading, and due to their inherently low baseline latency, they accommodate a significantly higher threshold without substantial performance impact. 
Reusable operators have the highest original kernel latency, but also the slowest relative latency growth with increased data loading, thus justifying an intermediate threshold.

\subsection{ Adaptive Fusion Strategy for OPG}
\label{sec:weight_mnager}

\noindent 
\textbf{Fusion-Compute-Memory Trade-off.}   
In various frameworks that execute DNN, operator fusion is often a key optimization~\cite{niu2021dnnfusion}. 
While operator fusion reduces kernel launch overhead and intermediate memory (aligning with our $M_{\mathrm{peak}}$ minimization objective in Section~\ref{sec:cp_sat_overlap}), overly aggressive fusion contradicts the layer-wise load capacity constraints ($C_\ell$ in Constraint C3).
\revisionno{Review-A\\(6)\\Review-C\\(2)\\Review-D\\(3)}{L}{0.0cm} 
\revisioncr{
Fusing multiple layers into a single kernel reduces the number of synchronization boundary where data loading could otherwise interleave with computation.
Specifically, fusing $k$ operators into a single kernel reduces the number of distinct execution stages available for scheduling data movement from $k$ down to 1, 
shrinking the combined load capacity to}  $C_{\mathrm{fused}} \approx \min(C_1, \dots, C_k)$ instead of $\sum C_i$. 
This directly impacts our CP-SAT solver's ability to distribute weight chunks across layers, as formalized by:

\[
\sum_{\ell \in L(w)} x_{w,\ell} \leq \sum_{\ell} C_\ell \quad \text{(Total chunk capacity)}
\]

Over-fusing shrinks the right-hand side, forcing more weights into preloading factor  ($W$) in our objective function. We quantify this factor n through \textit{fusion penalty scores} derived from the solver's residual capacity:

\[
\mathrm{Penalty}(v_{\mathrm{fused}}) = \underbrace{\lambda |W_{\mathrm{new}}|}_{\text{Preload cost}} + \underbrace{\mu \Delta z_w}_{\text{Distance penalty}} 
\]

where $v_{\mathrm{fused}}$ is a fused kernel node in DAG $G$, $W_{\mathrm{new}}$ are the weights forced into preload due to fusion and $\Delta z_w$ is the increased loading distance for the affected weights.

\begin{table}[t!]
    \centering
    \setlength{\tabcolsep}{3pt} 
    \caption{Operator classification and load capacity characteristics for representative operators: ``M.B.'' indicates memory bandwidth, ``L.C.'' denotes load capacity, and ``C.I.'' represents computational intensity.}
    \label{tab:operator_classification}
    \small
    \begin{tabular}{lccc}
        \toprule
        Operator Type & M.B. & L.C. Tolerance & C.I. \\
        \midrule
        Elemental ({\tt ReLU}, {\tt Add}) & Low & Medium & Low \\
        Reusable ({\tt Conv}, {\tt MatMul}) & Medium & High & High \\
        Hierarchical ({\tt LayerNorm}) & High & Low & Medium \\
        \bottomrule
    \end{tabular}
\end{table}

\compactparagraph{Adaptive Fusion Triggering.}
When the solver detects residual capacity violations, it triggers our fusion-aware adjustment protocol:  
\circled{1} \emph{Identify Critical Fusions:} Rank fused kernels by $\mathrm{Penalty}(v_{\mathrm{fused}})$ and select top candidates.  
\circled{2} \emph{Split Feasibility Check:} Verify if splitting $v_{\mathrm{fused}}$ into subkernels $\{v_1, v_2\}$ preserves DAG dependencies while satisfying:  
\[
C_{v_1} + C_{v_2} \geq (1 + \alpha) C_{v_{\mathrm{fused}}} \quad (\alpha > 0 \text{ is capacity gain threshold})
\]  
\circled{3} \emph{Iterative Refinement:} Update $G$ by replacing $v_{\mathrm{fused}}$ with $\{v_1, v_2\}$, then re-invoke the CP-SAT solver with adjusted $L(w)$ and $C_\ell$.  

\compactparagraph{Operator-Specific Splitting.}
Guided by our operator classification (Table~\ref{tab:operator_classification}), 
\revisionno{Review-C\\(3)}{L}{0.0cm} 
\revisioncr{we apply targeted splitting rules to restore scheduling flexibility. 
Here are two representative examples of the rules 
:
\circled{1} \emph{Reusable + Elemental Fusions:} Split into elemental (medium $C_\ell$) and reusable (high $C_\ell$) subkernels, e.g., Decompose ``{\tt MatMul}+{\tt Add}+{\tt GeLU}'' into ``{\tt MatMul}+{\tt Add}'' (reusable) and ``{\tt GeLU}'' (elemental), gaining $C_{\mathrm{total}} = C_{\mathrm{reusable}} + C_{\mathrm{elemental}}$.  }
\circled{2} \emph{Hierarchical Fusions:} Retain intact due to their low tolerance $C_\ell$ and synchronization requirements.

This strategy directly feeds into the layer capacity model (Section~\ref{sec:op_categories})--splitting a fused layer $v$ into $\{v_1, v_2\}$ increases schedulable load capacity by $C_{v_1} + C_{v_2} - C_v$, allowing more $x_{w,\ell}$ allocations and reducing preloads.

\subsection{Kernel Rewriting For Texture-optimized Layouts}
\label{sec:kernel_rewriting}

To further optimize GPU 2.5D texture memory performance, we design a kernel rewriting scheme that customizes data layout and removes control-flow divergence.
\hidetext{
\begin{figure}[t!]
    \centering
    \includegraphics[width=0.9\columnwidth]{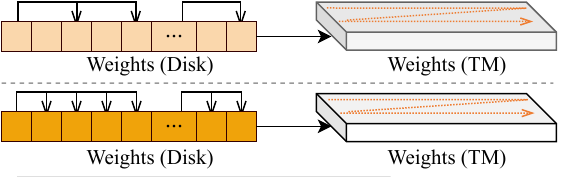}
    \caption{Read (on disk) and write (on texture memory) access patterns
    comparison before and after spatial data locality optimizations. \agrawal{I don't understand this figure}}
    \label{background:contiguous_read_write}
\end{figure}

\compactparagraph{Spatial Data Locality Optimization.}
GPU texture memory exhibits two-dimensional spatial locality, unlike the one-dimensional scheme typical of unified memory. Towards  exploiting this property, we reformat each weight tensor into a 2.5D micro-tiled layout offline on disk, ensuring that both disk reads and subsequent image writes remain contiguous. At runtime, we copy each tile directly from disk into the GPU's texture memory in a row-major order. The tiling strategy aligns tile boundaries with the hardware's inherent 2D caching structure, allowing disk coordinates to map naturally to coordinates in texture space. 

Figure~\ref{background:contiguous_read_write} illustrates that once the data is organized in this offline stage, the memory fetch operations  proceed without expensive  pointer arithmetic, and the texture cache achieves high hit rates during kernel execution. This off-line approach not only yields a more efficient  data flow, but also maximizes the performance benefits of 2.5D texture caching by unifying the read and write segments into a single, stride-friendly operation.
}

\compactparagraph{Pipelined Computation and Branch Divergence Elimination.}
\label{sec:pipelined_computation}
We introduce a kernel template  that interleaves arithmetic operation with texture-based weight loading and removes conditional branches in the process. 
A naive approach to interleaving compute and load often introduces conditional checks (for instance, to decide whether a thread should load data or perform computation), but such checks cause warp-level branch divergence and reduce SIMT efficiency. We avoid this by restructuring the loop to enforce a {\em uniform load–compute } schedule. 
Each iteration starts by prefetching the next tile's weights, then proceeds to compute the current tile without branch divergence.
When the pipeline is fully engaged, each iteration hides memory stalls from the newly issued loads behind the arithmetic of the prior iteration. The kernel concludes with a final loop pass that resolves any leftover arithmatic  operations.

\begin{figure}[t!]
    \centering
    \centerline{\includegraphics[width=0.95\linewidth]{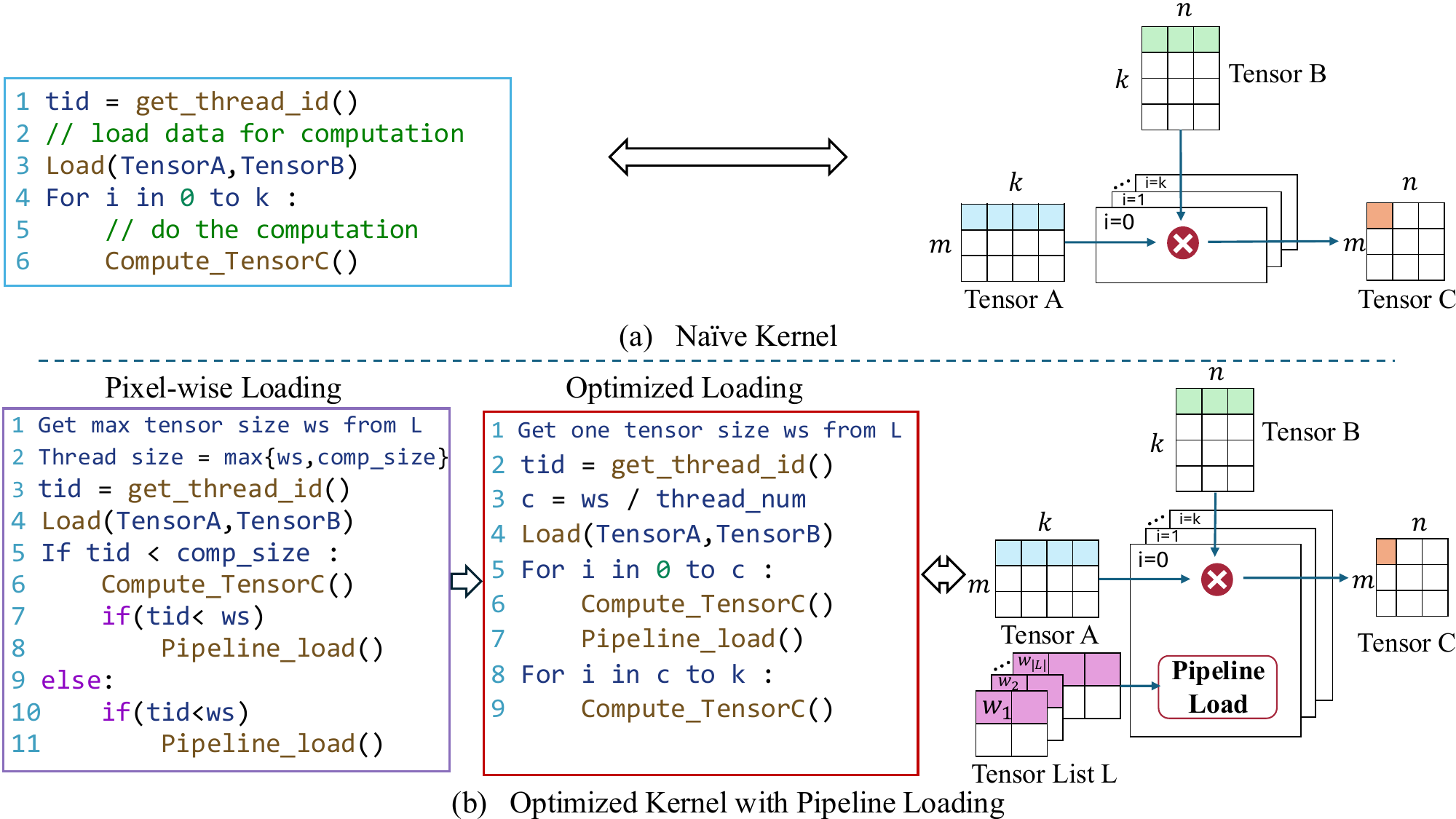}}
    \caption{Comparison of a naive Matrix Multiplication kernel with the
    rewritten kernel integrated with weight loading. }
    \label{kernel:rewriting}
\end{figure}

Figure~\ref{kernel:rewriting}(a) depicts a baseline matrix multiplication kernel where threads individually load TensorA and TensorB, then compute MAC (multiply-accumulate) outputs. In contrast, Figure~\ref{kernel:rewriting}(b) incorporates a fine-grained pipeline that alternates between partial compute and weight prefetches for the next block of Tensor~L. As soon as one block's MAC updates begin, the kernel schedules load instructions for the following block so that memory fetch latency overlaps with arithmetic on the current data.

This loop restructuring eliminates conditional logic and ensures that all threads follow the same execution path. Meanwhile, the pipeline mechanism matches the memory load phases with the MAC phases, preventing idle cycles. By aligning offline 2.5D tiling with a branch-free pipelined kernel, we achieve continuous data movement, effective latency hiding, and higher arithmetic utilization, thereby improving overall inference throughput on GPU texture architectures.

\revisionno{Review-A\\(1)}{L}{0.2cm}
\revisioncr{To reduce the engineering effort required to rewrite GPU kernels for each model, \projectname utilizes a template-based kernel rewriting approach~\cite{jinja}. Each DNN is represented as a computational graph made up of standard operators (e.g., {\tt MatMul}, {\tt Softmax}, {\tt Activation}). We create reusable kernel templates using the Jinja library~\cite{jinja}, which integrate the previously mentioned branch-free, pipelined weight loading directly into the computation. During deployment, the overlap plan specifies the loading timing for each operator, allowing the appropriate kernel template to be automatically instantiated from the computational graph without the need for model-specific code.}

\begin{table*}[t!]
    \centering
     \small
    \caption{Model characterization for evaluated models across different input types and tasks. ``Abbr.'' is used to refer to each model in Evaluation sections. ``\# MACs'' represents the number of multiply-accumulate operations. 
    \revisionno{Review-B\\(2)}{L}{0.0cm} \revisioncr{``\# Layers'' refers to the number of low-level operator nodes (e.g., {\tt MatMul}, {\tt Activation}) after graph lowering, instead of the high-level model blocks.} }
    \label{tab:model_comparison}
    \setlength{\tabcolsep}{8pt}
    \begin{tabular}{ll|cc|cccc}
    \toprule
    Model  & Abbr.     & Input Type & Model Task & \# Params (M) & \# MACs (G) & \# Layers \\ 
    \midrule
    GPTNeo-Small~\cite{gpt-neo} & GPTN-S    & Text  & NLP & 164   & 16   & 606   \\ 
    GPTNeo-1.3B~\cite{gpt-neo}  & GPTN-1.3B & Text  & NLP & 1,419 & 170  & 1,110 \\ 
    GPTNeo-2.7B~\cite{gpt-neo}  & GPTN-2.7B & Text  & NLP & 2,781 & 342  & 1,446 \\ 
    ResNet50~\cite{he2016deep}  & ResNet    & Image & Classification & 25.6 & 4.1 & 141 \\ 
    SegmentationAnything-2~\cite{ravi2024sam2}   & SAM-2   & Image & Segmentation       & 215 & 218  & 1,668             \\ 
    ViT~\cite{dosovitskiy2020vit}                & ViT     & Image & Classification     & 103 & 21   & 819              \\ 
    DeepViT~\cite{dosovitskiy2020vit}            & DeepViT & Image & Classification     & 204 & 42   & 1,395             \\ 
    StableDiffusion-UNet~\cite{StableDiffusion}  & SD-UNet & Image & Generation         & 860 & 78   & 1,271             \\ 
    Whisper-Medium~\cite{radford2023robust}      & Whisp-M & Audio & Speech Recognition & 356 & 55   & 2,026             \\ 
    DepthAnything-Small~\cite{depth_anything_v2} & DepA-S  & Video & Segmentation       & 24.3 & 14  & 1,108             \\ 
    DepthAnything-Large~\cite{depth_anything_v2} & DepA-L  & Video & Segmentation       & 333  & 180 & 2,007             \\ 
    \bottomrule
    \end{tabular}
\end{table*}

\section{Evaluation}  
\label{sec:evaluation}

\projectname is built on top of an existing end-to-end DNN execution framework on mobile GPUs named SmartMem~\cite{niu2024smartmem}.
This section evaluates \projectname against other state-of-the-art frameworks across multiple dimensions, including latency, memory usage, and energy consumption.
Specifically, the evaluation has the following 5 key objectives:
1) demonstrating \projectname's effectiveness in terms of latency and memory consumption compared to existing frameworks;
2) emphasizing the advantages of \projectname in  supporting large-scale models and enabling efficient multi-model execution, 
3) understanding the benefits of proposed optimizations through detailed breakdown studies;
4) evaluating the power and energy consumption with different models;
and 5) validating portability of \projectname across devices.

\begingroup
\begin{table*}[t!]
  \centering
  \setlength{\tabcolsep}{3pt} 
  \renewcommand{\arraystretch}{1.0}
  \caption{Overall latency consumption comparison. ``--'' means the model is not supported. Speedup refers to the latency ({\tt Init} + {\tt Exec}) of other frameworks over \projectname ({\tt Integrated}). \revisioncrs{Since \projectname considers the initialization and execution as a whole, we only report integrated latency. Solver is running offline, which is not included here.}} 
  \label{tab:overall_eval_final}
  \small
  \begin{tabular}{l|cc|cc|cc|cc|cc|cc|c|cc}
    \toprule
    \multirow{2}{*}{Model} & \multicolumn{2}{c|}{MNN (ms)} & \multicolumn{2}{c|}{NCNN (ms)} & \multicolumn{2}{c|}{TVM (ms)} & \multicolumn{2}{c|}{LiteRT (ms)} & \multicolumn{2}{c|}{ETorch (ms)} & \multicolumn{2}{c|}{SMem (ms)} & Ours  (ms) & \multirow{2}{*}{Speedup$^\dagger$} & \multirow{2}{*}{Speedup$^\ddagger$} \\
    ~ & Init & Exec & Init & Exec & Init & Exec & Init & Exec & Init & Exec & Init & Exec & Integrated & ~ & ~ \\ \midrule
    GPTN-S    & 3,529  & 337   & -- & --   & 5,832     & 621  & --   & --   & 277     & 5,869 & 4,757  & 59   & \textbf{577}   & \textbf{8.4$\times$}       & \textbf{9.3$\times$}       \\
    GPTN-1.3B & --     & --    & -- & --   & --        & --   & --   & --   & 5,178   & 515,291 & 48,109 & 501  & \textbf{3,086} & \textbf{15.8$\times$}       & \textbf{169$\times$}       \\
    GPTN-2.7B & --     & --    & -- & --   & --        & --   & --   & --   & --      & --     & --     & --   & \textbf{7,567} & \textbf{--}       & \textbf{--}       \\
    ResNet50  & 1,751  & 22    & 1,341 & 28 & 524       & 56   & 573  & 34   & 65      & 10,302 & 1,470  & 33   & \textbf{473}   & \textbf{3.2$\times$}       & \textbf{3.3$\times$}       \\
    SAM-2     & --     & --    & -- & --   & --        & --   & --   & --   & 1,178   & 857,752 & 9,983  & 826  & \textbf{1,267} & \textbf{8.5$\times$}       & \textbf{678$\times$}       \\
    ViT       & 2,550  & 476   & -- & --   & 3,527     & 841  & 711  & 91   & 90      & 6,671 & 3,675  & 73   & \textbf{347}   & \textbf{10.8$\times$}       & \textbf{8.4$\times$}       \\
    DeepViT   & 4,345  & 883   & -- & --   & 6,243     & 1,665& 1,013& 254  & 298     & 60,656 & 7,699  & 190  & \textbf{785}   & \textbf{10.0$\times$}       & \textbf{9.6$\times$}       \\
    SD-UNet   & 21,747 & 1,647 & -- & --   & --        & --   & --   & --   & 7,692   & 1,056,869 & 29,588 & 312  & \textbf{3,212} & \textbf{9.3$\times$}       & \textbf{49.1$\times$}       \\
    Whisper-M & 6,143  & 1,343 & -- & --   & 7,256     & 2,157& --   & --   & --      & --     & 15,066 & 336  & \textbf{1,565} & \textbf{9.8$\times$}       & \textbf{5.4$\times$}       \\
    DepthA-S  & 2,492  & 588   & -- & --   & 2,012     & 487  & --   & --   & --      & --     & 2,200  & 71   & \textbf{496}   & \textbf{4.6$\times$}       & \textbf{5.6$\times$}       \\
    DepthA-L  & 6,267  & 1,784 & -- & --   & 6,988     & 1,917& --   & --   & --      & --     & 18,567 & 807  & \textbf{1,382} & \textbf{14.0$\times$}       & \textbf{6.1$\times$}       \\\midrule
    Geo-Mean  & \multicolumn{2}{c|}{\bf 6.1$\times$}  & \multicolumn{2}{c|} {\bf 2.9$\times$}  & \multicolumn{2}{c|}{\bf 6.2$\times$}  & \multicolumn{2}{c|}{\bf 1.7$\times$}  & \multicolumn{2}{c|}{\bf 75$\times$}  & \multicolumn{2}{c|}{\bf 8.6$\times$}  & \textbf{1.0$\times$}  & N/A  & N/A  \\ 
    \bottomrule
    \multicolumn{16}{l}{$^\dagger$: geo-mean speedup over SmartMem (a precursor research prototype). }\\
    \multicolumn{16}{l}{$^\ddagger$: geo-mean speedup over other frameworks (all commercial/product-level systems).}\\
  \end{tabular}
  \vspace{-0.6em}
\end{table*}
\endgroup

\subsection{Experimental Setup}

\compactparagraph{Models, Datasets, and Accuracy.}
To evaluate \projectname, we conduct experiments on 11 models that cover six representative mobile application tasks: NLP, image classification, image segmentation, image generation, speech recognition, and video segmentation. 
A detailed comparison of the characteristics of the model is provided in Table~\ref{tab:model_comparison}, including the target task, the number of parameters, the total number of layers, and the number of multiply-accumulate operations (MACs).
The abbreviation column is used to refer to each model in the following sections.
Depending on the target
task, the models are trained on the ImageNet~\cite{deng2009imagenet} dataset (for image classification and segmentation), 
the LibriSpeech~\cite{panayotov2015librispeech} dataset (for speech recognition), the datasets mentioned in the original DepthAnything paper \cite{depth_anything_v2} (for image segmentation), the Pile~\cite{gao2020pile} dataset from EleutherAI (for NLP task) and the LAION-5B~\cite{schuhmann2022laion} dataset (for image generation).
Since all frameworks use the same pre-trained model and repeated trials confirm the same accuracy, we only report inference performance on mobile devices.  

\compactparagraph{Baselines.}
We compare \projectname with the state-of-the-art mobile DNN frameworks -- LiteRT~\cite{TensorFlow-Lite} (TFLite a32930a7), ExecuTorch~\cite{executorch} (ETorch 967e3b99), MNN~\cite{Ali-MNN} (daa62c77), NCNN~\cite{Ni_ncnn_2017} (d395000e), and TVM~\cite{chen2018tvm} (89f9573d). 
We do not include frameworks like FlexNN~\cite{li2024flexnn} and llama.cpp~\cite{llamacpp} in this comparison because they do not currently support the models tested on mobile GPUs.
Our comparison also included the precursor research prototype, SmartMem~\cite{niu2024smartmem}. 

\compactparagraph{Target Mobile Devices.}
Our evaluation considers four smartphones that span a broad range of computational resources: the OnePlus 12, OnePlus 11, Xiaomi Mi 6, and Google Pixel 8. The OnePlus 12 integrates an Adreno 750 GPU and 16 GB of RAM; the Google Pixel 8 incorporates a Mali-G715 MP7 GPU and 8 GB of RAM; the OnePlus 11 employs an Adreno 740 GPU and 16 GB of RAM; and the Xiaomi Mi 6 combines an Adreno 540 GPU and 6 GB of RAM; 
Unless otherwise specified, all reported results correspond to the OnePlus 12, while Section \ref{sec:eva-portability} presents a detailed portability analysis of \projectname on Google Pixel 8, OnePlus 11, and Xiaomi Mi 6.


\compactparagraph{Experimental Configurations.}
We utilize the 16-bit and 32-bit\footnote{32-bit results, showing similar trends to 16-bit, are in the appendix.} floating-point data types for GPU execution,
as they are  supported by all frameworks. 
Models using lower precision, such as quantized models, are excluded from our evaluation due to the lack of support for them in other frameworks (at least for the models used in our study and on mobile GPUs). 
All experiments use a batch size of 1.  
To optimize performance, we utilize the auto-tuning features of MNN, NCNN, TVM, LiteRT, ExecuTorch, and SmartMem.  
Each experiment is run 50 times; only the average results are reported since variance is negligible.

\begin{table}[t!]
  \centering
  \small
  \setlength{\tabcolsep}{1.0pt}
  \renewcommand{\arraystretch}{1.0}
  \caption{Overall memory consumption comparison. ``--'' means the model is not
  supported. ``Mem-ReDT'' refers to the memory reduction over SmartMem. }
  \label{tab:overall_eval_mem}
  \begin{tabular}{l|ccccccc|c}
    \toprule \multirow{2}{*}{ Model} & \multicolumn{7}{c|}{ Average Memory (MB)} & { Mem-}          \\
    ~                                & MNN                                       & NCNN            & TVM             & LiteRT          & ETorch          & SMem            & {\bf Ours}      & {\bf ReDT}        \\
    \hline
    GPTN-S                           & 610                                       & --              & 2,300           & --              & 702             & 541             & 260             & {\bf 2.1$\times$} \\
    GPTN-1.3B                        & --                                        & --              & --              & --              & 2,600           & 2,667           & 554             & {\bf 4.8$\times$} \\
    GPTN-2.7B                        & --                                        & --              & --              & --              & --              & --              & 1,132           & --                \\
    ResNet50                         & 149                                       & 165             & 789             & 331             & 129             & 140             & 83              & {\bf 1.7$\times$} \\
    SAM-2                            & --                                        & --              & --              & --              & --              & 896             & 150             & {\bf 6.0$\times$} \\
    ViT                              & 369                                       & --              & 801             & 711             & 375             & 390             & 83              & {\bf 4.7$\times$} \\
    DeepViT                          & 824                                       & --              & 3,072           & 2,355           & 1,228           & 826             & 165             & {\bf 5.0$\times$} \\
    SD-UNet                          & 1,800                                     & --              & --              & --              & 1,792           & 2,100           & 838             & {\bf 2.5$\times$} \\
    Whisper-M                        & 1,650                                     & --              & 1,638           & --              & --              & 1,433           & 240             & {\bf 6.0$\times$} \\
    DepthA-S                         & 148                                       & --              & 461             & --              & --              & 150             & 86              & {\bf 1.7$\times$} \\
    DepthA-L                         & 1,230                                     & --              & 1,260           & --              & --              & 1,200           & 246             & {\bf 4.9$\times$} \\
    \midrule Geo-Mean                & \bf 3.2$\times$                           & \bf 2.0$\times$ & \bf 8.4$\times$ & \bf 7.9$\times$ & \bf 3.4$\times$ & \bf 3.5$\times$ & \bf 1.0$\times$ & N/A               \\
    \bottomrule
  \end{tabular}
\end{table}

\subsection{Overall Performance Evaluation}
\label{sec:overall_performance} 

{\noindent \bf End-to-end latency comparison.} 
Table~\ref{tab:overall_eval_final} presents an end-to-end comparison of the overall execution latency across multiple frameworks on our target mobile GPU, 
\revisionno{Review-C\\(4)}{L}{0.0cm} 
\revisioncr{For all other frameworks that preload the data, we report  model initialization time (i.e., cold-start scenario, weight preload and transformation from unified memory to texture memory) and execution time separately. For \projectname, as these two phases are integrated,  only 
a single time is reported. }
The results of transformer-based models for NCNN are omitted due to the lack of operator supports (e.g., LayerNorm) for these models on mobile GPUs.
SmartMem supports most of the models among the baseline frameworks, so we list the speedup of \projectname over SmartMem and other frameworks separately (as indicated in the last two columns).
Overall,  when comparing integrated times for all frameworks, \projectname outperforms all other frameworks, achieving an average speedup of 6.1$\times$, 2.9$\times$, 6.2$\times$, 1.7$\times$, 75.0$\times$, and 8.6$\times$ over MNN, NCNN, TVM, LiteRT, ExecuTorch, and SmartMem, respectively.
Compared to ExecuTorch, \projectname achieves a significant speedup because ExecuTorch lacks effective GPU-specific optimizations, which prevents it from fully utilizing hierarchical GPU memory and leads to inefficient resource utilization.
None of the other frameworks supports GPTN-2.7B, as it is a large-scale model with 2.7 billion parameters, exceeding the capacity of these frameworks. 
It is worth pointing out that \projectname is particularly effective for large-scale models like GPT-Neo-1.3B , DeepViT, and SD-UNet, where it achieves speedups of 15.8$\times$, 10.0$\times$, and 9.3$\times$, 
respectively, over SmartMem. For these three models, speedup  numbers over  other frameworks are 168.7$\times$, 9.6$\times$, and 49.1$\times$, respectively.
The reason for particularly stronger performance is that \projectname's overlapping and fusion strategy  reduces the overhead of preload and data transformation. 
\projectname does have a higher integrated (Init and Execution) time as compared to 
SmartMem's  inference-only time. 
\revisionno{Review-C\\(4)}{L}{0.0cm} 
\revisioncr{Thus, for models that SmartMem supports, it can be relatively faster in a warm-start setting after 3--12 consecutive inference tasks using the same model.} 




\noindent\textbf{End-to-end memory comparison.} 
To evaluate the memory efficiency of \projectname, we compare its average memory consumption with existing frameworks, i.e. each of  MNN, NCNN, TVM, LiteRT, ExecuTorch, and SmartMem used in previous experiment.  Table~\ref{tab:overall_eval_mem} reports the memory usage for various models, along with the memory reduction factor (Mem-ReDT) relative to SmartMem. Models that are not supported by a given framework are marked with ``–''. 
Overall, \projectname achieves an average memory reduction of 3.2$\times$, 2.0$\times$, 8.4$\times$, 7.9$\times$, 3.4$\times$, and 3.5$\times$ over MNN, NCNN, TVM, LiteRT, ExecuTorch, and SmartMem, respectively.
Convolution-based models like SD-UNet (2.5$\times$ savings) and DepthAnything-Small (1.7$\times$ savings) show lower memory reductions because they require convolution weight transformations (e.g., Winograd transforms) that temporarily increase memory usage during execution.
Even with this additional overhead, \projectname's overlap plan generation avoids loading and retaining all weights at once, thus offering significant memory benefits on devices with limited RAM.

\begin{figure}[t!]
  \centering
  \centerline{\includegraphics[ height=0.8\columnwidth]{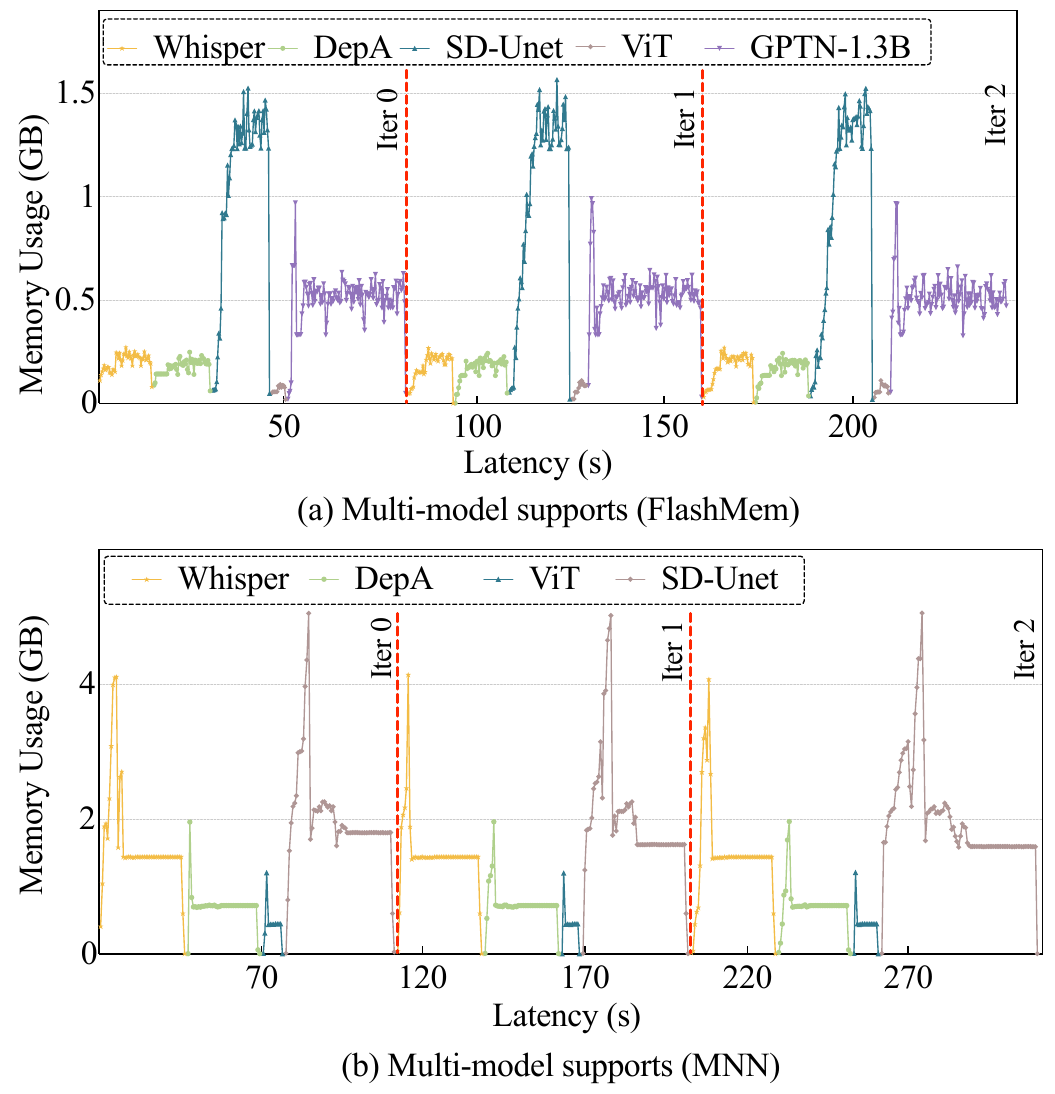}}
  \caption{\revisionno{Review-B\\(4)}{R}{0.0cm}\revisioncr{Multi-model supports on \projectname (a) and MNN (b), each model runs for 10 iterations interleaved.}}
  
  \label{eval:corunning}
\end{figure}

\subsection{Multi-model Support Evaluation}
We further evaluate the effectiveness of \projectname in managing memory consumption under multi-model workloads on mobile GPUs. Figures~\ref{eval:corunning} illustrate the memory usage patterns of four representative models running sequentially in a random order.
\revisionno{Review-B\\(3)}{R}{0.0cm}
\revisioncr{FlashMem uses a \revisioncrs{manually} selected 1.5 GB constraint to prioritize the  latency.}
MNN exhibits significant peak memory usage during the initialization of each model, reflecting redundant data transformations and multiple copies of weights in memory. 
On the other hand, \projectname drastically reduces peak memory by generating an efficient overlap plan for fine-grained weight loading and  kernel computations. 
This reduction becomes more substantial for larger models (e.g., SD-UNet), due to their higher initial memory overhead.

\subsection{Detailed Analysis} 

\begin{figure}[t!]
  \centering
  \centerline{\includegraphics[width=\columnwidth]{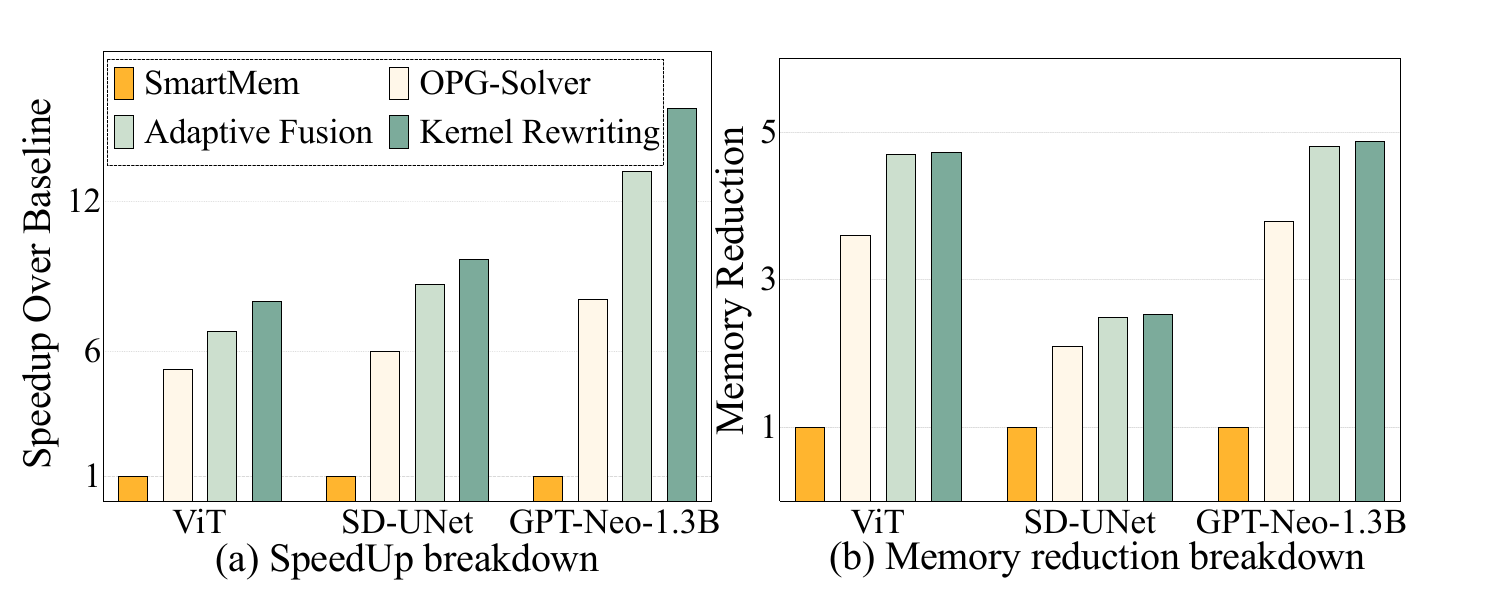}}
  \caption{Performance breakdown analysis: speedup (a) and memory reduction (b) over baseline (SmartMem).}
  \label{eval:breakdown}
\end{figure}

\compactparagraph{Latency and Memory Breakdown Analysis.}
We analyze the impact of the key optimizations of \projectname on inference latency. Figure~\ref{eval:breakdown} (a) presents the breakdown of end-to-end inference latency improvements for three representative models: ViT, SD-UNet, and GPT-Neo-1.3B. Due to space limitations, we omit results for other models, which exhibit similar trends. 
We evaluated the incremental contributions of \emph{OPG-Solver}, \emph{Adaptive Fusion}, and \emph{Kernel Rewriting} in the baseline framework (SmartMem). \emph{OPG-Solver} achieves speedups ranging from 5.3$\times$ to 8.1$\times$, while \emph{Adaptive Fusion} and \emph{Kernel Rewriting} provide additional improvements of 1.5$\times$ to 5.1$\times$ and 1.0$\times$ to 2.55$\times$, respectively. 
Larger models, such as GPT-Neo-1.3B and SD-UNet, benefit more compared to smaller models like ViT, as they are more constrained by bandwidth-bound data loading and preparation.


In addition to latency improvements, we evaluate the impact of \projectname's key optimizations on average memory consumption. As shown in Figure~\ref{eval:breakdown} (b), \emph{OPG-Solver} achieves memory reductions ranging from 2.1$\times$ to 3.8$\times$, while \emph{Adaptive Fusion} and \emph{Kernel Rewriting} further contribute reductions of 1.1$\times$ to 1.4$\times$ and 1$\times$ to 1.1$\times$, respectively. Kernel rewriting primarily contributes to latency reduction, with fewer gains in memory efficiency. 
The overall improvements in latency and memory usage resulting from these optimizations can be attributed to three key factors: deferring weight loading until necessary, 
reducing redundant data copies, 
and overlapping memory transfers with computation, collectively minimizing inference latency and memory 
footprints. Additionally, our insight is that for convolution-based models (i.e., SD-UNet, DepthAnything), memory and latency reductions are generally less significant because convolutional weight transformations required for efficient computation cannot be overlapped. In contrast, for transformer-based models (i.e., others except for SD-UNet and DepthAnything), larger model sizes yield more substantial benefits in both latency and memory usage reductions.


\begin{figure}[t!]
\centering
\centerline{\includegraphics[width=\columnwidth]{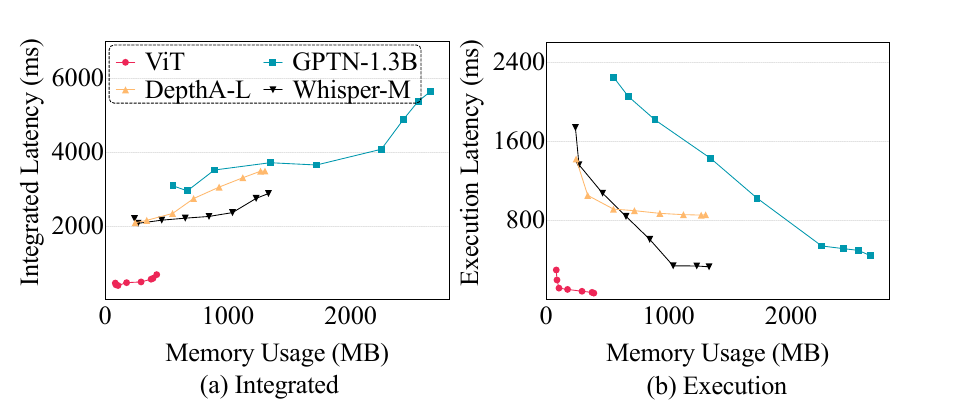}}
\caption{Trade-off between memory usage and inference latency across different models, showing integrated latency (a) and execution latency (b). \revisioncrs{As more weights are preloaded before inference, we will have a lower execution latency. However, initialization latency becomes the dominant factor of the integrated latency, which introduces higher total latency.}}
\label{eval:preload_ratio}
\end{figure}

\compactparagraph{Latency and Memory Tradeoff.} 
Figure~\ref{eval:preload_ratio} provides a configurable trade-off between memory usage and inference latency, 
controlled by parameters including the memory peak constraint ($M_{peak}$), 
the memory usage penalty ($\mu$), and the preload relaxation factor ($\lambda$). 
Larger models are more sensitive to the preload ratio due to higher memory pressure and I/O overhead. 
For example, GPT-Neo-1.3B sees a sharp latency reduction  as preloading increases,
since loading only a small portion of its weight leads to frequent transfers,
resulting in significant I/O overhead and high inference delays.

A key observation from the figure is that the overlap of an average of 49.3\% of the weights results in negligible additional latency overhead compared to full loading, 
while significantly reducing memory usage. 
This indicates that an optimal trade-off is achievable by partially preloading weights, 
preventing excessive memory consumption without incurring high runtime loading costs.
A higher preload ratio can be achieved by increasing $\lambda$ and setting relatively lower values for $M_{peak}$ and $\mu$. 
For additional flexibility, weights can also be explicitly specified by directly adding their names to the preload list $|W|$.

\begin{figure}[t!]
  \centering
  \centerline{\includegraphics[width=0.865\columnwidth]{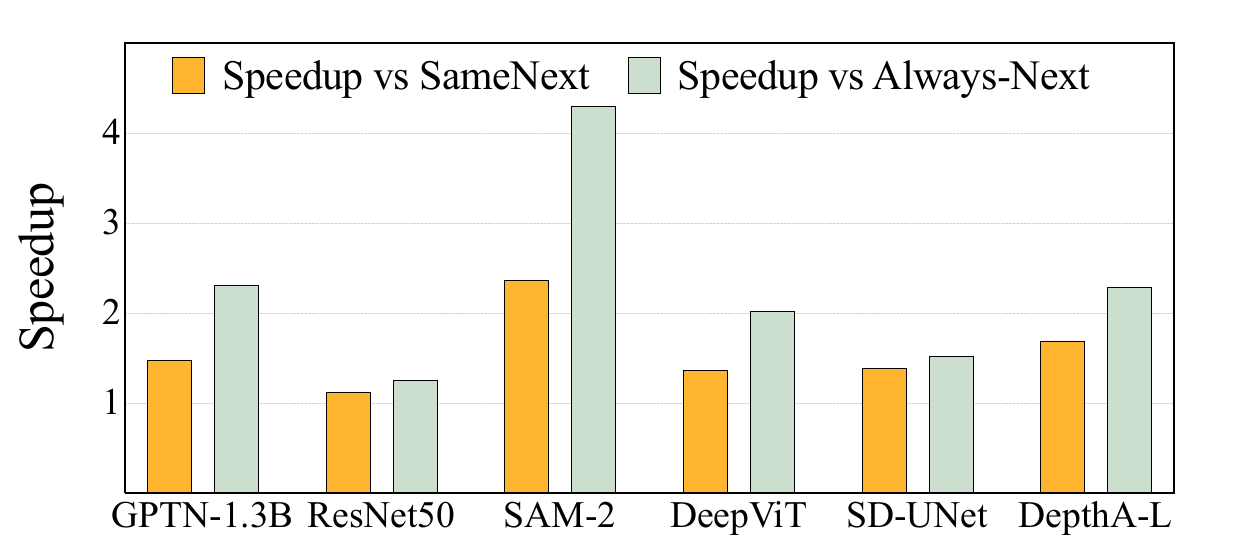}}
  \caption{Latency comparison among Ours (\projectname), Same-Op-Type Prefetching (SameNext), and Always-Next Loading.}
  \label{fig:overlap-strategies}
\end{figure}

\compactparagraph{Comparison with Naive  Overlap Strategies.} 
To evaluate the effectiveness of our overlap policy, we conduct additional experiments comparing \projectname to two baselines, as shown in Figure~\ref{fig:overlap-strategies}.
The first, {\em Always-Next Loading}, prefetches the weight for the next layer, which can cause GPU transformation step  to lag behind the disk loading. We can see from Figure~\ref{fig:overlap-strategies} 
that this results  in up to 4.3$\times$ slower performance than \projectname. 
The second method, {\em Same-Op-Type Prefetching}, preloads weights only from layers of the same type. 
Thus, compared to the first schemes, this method partially considers loading 
capacity. However, the computation and data movement continues to be  imbalanced across the model, resulting in execution  up to 2.4$\times$ slower.
This highlights the importance of adopting a balanced strategy for prefetching and loading in \projectname.

\begin{table}[t!]
  \centering
  \small
  \setlength{\tabcolsep}{6pt}
  \caption{Comparison of average power and energy consumption: energy is calculated by multiplying the average power by the latency of each iteration. \revisioncrs{Power was measured by reading the system power usage over time.}}
  \begin{tabular}{l|cc|cc}
    \toprule \multirow{2}{*}{Model} & \multicolumn{2}{c|}{Power Consumption} & \multicolumn{2}{c}{Energy Consumption} \\
    ~                               & {DeepViT}                              & {SD-UNet}                             & {DeepViT} & {SD-UNet} \\
    \midrule MNN                    & 6.3W                                   & 4.8W                                  & 33.1J     & 95.2J     \\
    LiteRT                          & 6.4W                                   & --                                    & 51.3J     & --        \\
    ETorch                          & 3.6W                                   & --                                    & 130.5J    & --        \\
    SmartMem                        & 5.2W                                   & 4.5W                                  & 41.0J     & 134.5J    \\
    Ours                            & 5.7W                                   & 5.6W                                  & 4.5J      & 17.9J     \\
    \bottomrule
  \end{tabular}
  \label{eval:power_comparison}
\end{table}

\subsection{Power and Energy Consumption}
\label{tab:eva-power-consumption}
Table~\ref{eval:power_comparison} presents power consumption against other frameworks in two representative models: DeepViT, and SD-UNet. 
For DeepViT, \projectname achieves a power consumption comparable to or slightly higher than Executorch and SmartMem, while less than MNN and LiteRT.
Our power consumption higher than SmartMem due to the extra data movement between disk and GPU memory, and it is higher than ExecuTorch since we have a better GPU utilization.
As for energy consumption, \projectname saves 83\%, 91\%, 96\% and 87\% compared to MNN, LiteRT, ExecuTorch, and SmartMem, respectively.
The benefits result from lower execution latency for model initialization and inference. 
\revisionno{Review-A\\(2)}{L}{0.0cm} 
\revisioncr{A quantitative analysis of the trade-off between memory savings and power/energy consumption 
is beyond the scope of this paper and could  be a promising future research direction.}

\begin{figure}[t!]
    \centering\centerline{\includegraphics[width=0.8\columnwidth]{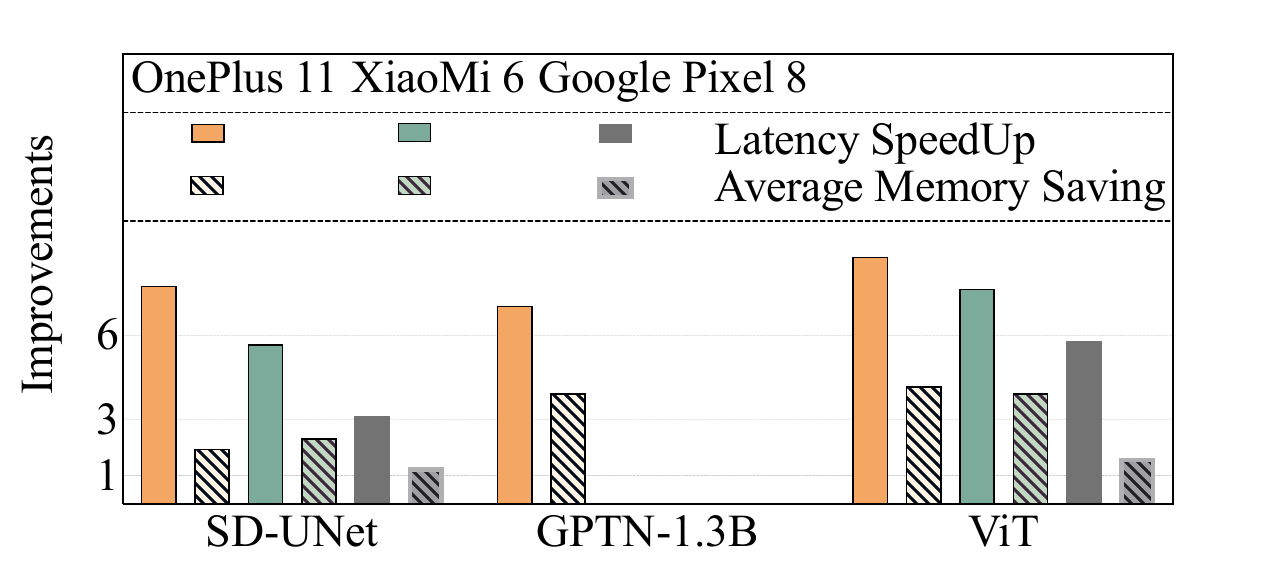}}
    \caption{Comparison of memory usage and latency across three devices using SmartMem versus \projectname. Empty bars indicate the device ran out of memory during initialization. 
    }
    \vspace{-0.5em}
    \label{eval:portability}
\end{figure}

\subsection{Portability Evaluation}
\label{sec:eva-portability}
This section presents the results on different mobile GPUs with varied computational and memory capacities, specifically the Oneplus 11, Xiaomi Mi6, and Google Pixel 8 as illustrated in Figure~\ref{eval:portability}.
\projectname achieves consistent performance improvements over SmartMem despite limited resources. 
GPTN-1.3B is not supported in Xiaomi Mi 6 and Google Pixel 8, due to its limited 6GB memory capacity and high memory usage during initialization. Note that all three models can be run on these devices by using \projectname.
Figure~\ref{eval:portability} shows that \projectname exhibits 
good portability across various devices, and it is even more resilient to 
limited hardware resources due to the reduced memory requirements.

\section{Related Work}

\compactparagraph{Efficient Memory Management for DNN Acceleration.}
Memory efficiency is a key challenge in accelerating deep neural networks (DNNs) on resource-constrained devices~\cite{huynh2017deepmon,ye2024asteroid,wang2024swapnet,tan2024cocco,li2021chimera,yao2017deepsense,dreuning2024capslog,tang2024delta,zhou2024deeptm,jeong2022band}. 
Asteroid \cite{ye2024asteroid} employs hybrid pipeline parallelism to enable collaborative training across heterogeneous devices. 
Chimera \cite{li2021chimera} leverages bidirectional pipelining to improve training efficiency and resource utilization for large-scale networks.
Checkmate \cite{jain2020checkmate} presents an approach to minimize memory footprint by optimally rematerializing tensors, thus reducing redundant data storage.
In the inference domain, SwapNet \cite{wang2024swapnet} partitions the models into blocks and swaps them dynamically between disk and memory, allowing execution beyond the limits of hardware memory. 
Cocco \cite{tan2024cocco} further optimizes hardware mapping by reducing communication overhead and memory access using a graph-partitioning approach.
vLLM \cite{pageattention} introduces a paging strategy to dynamically manage attention weights during inference, reducing memory usage in LLM serving.

FlexNN \cite{li2024flexnn} introduces adaptive memory management with slicing-loading-computing planning to optimize resource allocation on mobile CPUs majorly benefiting for convolutional models.
Fu  {\em et al.} ~\cite{fu2024heterogeneous} propose heterogeneous memory integration for NLP tasks on customized accelerators. 
Pantheon~\cite{han2024pantheon}  designs a preemptive scheduling system that enables real-time tasks to preempt each other.
Their focus is not on texture memory hierarchy issues, and the work does not support very large models on mobile GPUs.

\compactparagraph{End-to-end DNN Executions on Mobile Platforms.}
With the growing demand for AI applications on mobile devices, optimizing the execution of DNN for resource-constrained platforms has become a critical research area. 
Early works~\cite{han2016mcdnn,lane2016deepx,huynh2017deepmon,yao2017deepsense,xu2018deepcache,niu2021dnnfusion} primarily focused on optimizing the execution of  DNNs that were  smaller in size and less memory-intensive. 
Modern product-level frameworks such as LiteRT (formerly TensorFlow Lite), PyTorch Mobile, TVM, and MNN enable efficient mobile inference across various model architectures.
Recent works, such as PowerInfer~\cite{xue2024powerinfer,song2024powerinfer} and vAttention~\cite{prabhu2025vattention}, introduces memory management techniques for running large language models while reducing physical memory fragmentation.
However, they require loading entire models into memory, resulting in high memory consumption.
This limitation prevents the efficient execution of large models and 
multi-DNN execution scenarios, which \projectname targets. 

Recent studies, including llama.cpp~\cite{llamacpp}, llamafile~\cite{llamafile}, ollama~\cite{ollama}, Xu et al.~\cite{xu2025fast}, and MLC-LLM~\cite{mlc-llm}, have focused on optimizing large-scale LLMs execution on mobile devices, primarily enhancing inference efficiency for CPU and/or specialized accelerators. 
These approaches do not support memory offloading techniques necessary for handling large models under strict memory constraints. Moreover, they lack optimizations for accelerating DNNs on mobile GPUs with 
emphasis on texture memory related issues.

\section{Conclusion}
\label{sec:conclusion}

This paper has presented  \projectname, an efficient framework for overlapping data movement and computation to optimize on-device DNN execution under memory constraints, with the goal 
of supporting large models and/or workloads involving multiple models. 
By statically scheduling weight loading and balancing computation with data movement at a fine granularity, \projectname reduces peak memory usage while improving execution efficiency. 
It  optimizes for GPU texture memory (hierarchy),  resulting in low transformation overheads and 
maximizes   GPU resource utilization through rewriting of kernels and other optimizations. 
Experimental evaluation  across diverse mobile platforms has demonstrated  that 1) \projectname can enable execution of models like GPTN-2.7B that currently no other Mobile DNN framework can  support, 
2) across different modern DNN models, reduces execution latency and memory requirements over state-of-the-art frameworks,  and 3) can support workloads 
involving multiple DNN models effectively. 
\revisionno{Review-C\\(1)}{L}{0.0cm} 
\revisioncr{In the future, we plan to extend this methodology to datacenter settings, and explore the potential benefits and challenges in large scale deployments.}


\begin{acks}
\revisioncr{
The authors want to extend their appreciation to the anonymous reviewers and shepherd, Martin Maas, for their valuable and thorough feedback. 
All of these constructive suggestions have greatly contributed to enhancing this paper. 
This work was supported in part by the National Science Foundation (NSF) under the awards of 
CCF-2428108, 
CCF-2333895,
OAC-2403090 
and CSR-2341378.
Any errors and opinions are not those of the NSF and are attributable solely to the author(s). 
}
\end{acks}

\bibliographystyle{ACM-Reference-Format}
\balance
\bibliography{acmart.bib}

\end{document}